\documentclass[epj]{svjour}          
\usepackage{slashed}
\usepackage{graphicx,amssymb,amsmath}
\usepackage{amsfonts}
\usepackage{nicefrac}
\usepackage{widetext}
\usepackage[usenames]{color}
\usepackage[colorlinks=true,linkcolor=blue,citecolor=blue,urlcolor=blue]{
hyperref}
\usepackage{cite}

\newcommand{\beq}{\begin{equation}}
\newcommand{\eeq}{\end{equation}}

\newcommand{\flaligne}[1]{\begin{flalign} #1 \end{flalign}}

\newcommand{\bracket}[1]{\left( #1 \right)}

\begin{document}
\title{On the effect of resonances in the quark-photon vertex}

\author{\'Angel S. Miramontes \and H\`elios Sanchis-Alepuz\inst{1}}                     
\institute{Institute of Physics, University of Graz, NAWI Graz, 8010 Graz, Austria}
\date{Received: date / Revised version: date}

\abstract{
A calculation of hadronic timelike form factors in the Poincar\'e-covariant Bethe-Salpeter formalism necessitates knowing the analytic structure of the non-perturbative quark-photon vertex in the context of the Poincar\'e-covariant Bethe-Salpeter formalism. We include, in the interaction between quark and antiquark, the possibility of non-valence effects by introducing pions as explicit degrees of freedom. These encode the presence of intermediate resonances in the Bethe-Salpeter interaction kernel. We calculate the vertex for real as well as complex photon momentum. We show how the vertex reflects now the correct physical picture, with the rho resonance appearing as a pole in the complex momentum plane. A multiparticle branch cut for values of the photon momentum from $-4m_\pi^2$ to $-\infty$ develops. This calculation represents an essential step towards the calculation of timelike form factors in the Bethe-Salpeter approach.  
} 

\maketitle

\section{Introduction}\label{sec:intro}

The quark-photon vertex describes the interaction of quarks with photons in quantum field theory. It is, therefore, a crucial ingredient in the study of the electromagnetic interaction of hadrons. For example, the electroproduction of hadrons off nucleons, which is the main experimental technique for the study of nucleon resonances (see, e.g. \cite{Aznauryan:2012ba}), is described microscopically (in the one-photon approximation) by the exchange of a spacelike virtual photon that couples to the quarks forming the nucleon. The coupling of quarks to timelike photons determines instead the processes of hadronic particle-antiparticle creation or annihilation, as studied experimentally at BES-III \cite{Asner:2008nq} and in the future at PANDA \cite{Wiedner:2011mf}. Those couplings are described, respectively, by the spacelike and timelike form factors of hadrons \cite{Denig:2012by,Pacetti:2015iqa,Punjabi:2015bba}.

It is by now well known that the strong interactions among quarks generates a structure of the quark-photon vertex much richer than its tree-level component $\gamma^\mu$ (see, e.g. \cite{Ball:1980ay,Frank:1994mf,Maris:1999bh,Chang:2010hb,Eichmann:2014qva,Tang:2019zbk}). In particular, for timelike photon momentum, the quark-photon vertex must reflect the full excitation spectrum of quantum chromodynamics (QCD) in the vector-meson channel, a fact which is at the heart of the phenomenological success of vector-meson dominance models \cite{Sakurai,Bauer:1977iq} (see also \cite{Leupold:2012qn}). The details of such a rich structure of the vertex are, however, not precisely known since they are generated by non-perturbative QCD effects.

The study of non-perturbative phenomena within continuum QCD can be approached using Dyson-Schwinger (DSE) and Bethe-Salpeter (BSE) equations. DSEs are non-linear integral equations describing the Green's functions (GFs) of the theory and BSEs are linear integral equations for bound states. Even though a full non-perturbative treatment of the quark-photon vertex would require to solve the corresponding DSE, if one is interested in QCD effects only, these can be studied with equations simpler than DSEs, namely inhomogeneous BSEs. The combination of DSEs and (homeogeneous and inhomogeneous) BSEs has been extensively and successfully used to study hadron phenomenology (see, e.g. \cite{Cloet:2013jya,Eichmann:2016yit,Huber:2018ned} and references therein). 

The complexity of non-perturbative calculations entails that nearly always some sort of approximation or simplification is necessary. In the context of DSEs and BSEs, it is necessary to truncate the infinite system of coupled DSEs that describe the theory and the infinite number of interaction terms in a BSE, as described below. Truncations of ever increasing sophistication that perform well phenomenologically have been developed over the years (see e.g. \cite{Sanchis-Alepuz:2014wea,Williams:2015cvx,Eichmann:2016yit,Qin:2019hgk,Tang:2019zbk} and references therein). To the best of our knowledge, however, the rainbow-ladder (RL) truncation of the BSE interaction kernel described below is the most sophisticated truncation used so far in the calculation of hadron form factors. Even though it performs remarkably well on the spacelike momentum region \cite{Maris:2000sk,Maris:2002mz,Bhagwat:2006pu,Cloet:2008re,Nicmorus:2010sd,Eichmann:2011vu,Eichmann:2011pv,Eichmann:2011aa,Sanchis-Alepuz:2013iia,Chang:2013nia,Segovia:2014aza,Sanchis-Alepuz:2015fcg,Segovia:2016zyc,Sanchis-Alepuz:2017mir,Chen:2018nsg,Chen:2018rwz,Ding:2018xwy}, the trend in all cases is that the RL truncation is insufficient to describe the behaviour of form factors at low photon momentum. The reason for that is usually attributed to RL calculations lacking so-called meson-cloud effects on form factors, which stem from the photon coupling to non-valence quarks inside the hadron (sea quarks).

In this work we study an extension of the RL truncation which encodes, to some extent, the above-mentioned non-valence quark effects on the BSE interaction kernel. The truncation studied herein was put forward in \cite{Fischer:2007ze,Fischer:2008sp} and keeps the resonant contributions of non-valence terms only, describing them in terms of explicit pionic degrees of freedom. In particular, the new kernel includes a virtual decay channel of e.g. a vector meson into two pions. We will show that this kernel generates the correct physical picture of the quark-photon vertex on the timelike momentum side, as mentioned above. Moreover, even though beyond the scope of this paper, it is reasonable to speculate that a calculation of form factors with the kernel used in this work must, to some extent, alleviate the problem of missing meson-cloud effects.

This work is organised as follows. In Sec.~\ref{sec:Formalism} we briefly describe the basic elements of the DSE/BSE formalism necessary for the present calculation, including a discussion of the role of symmetries in the choice of the BSE interaction kernel. Non-standard numerical techniques \cite{Windisch:2012sz,Windisch:2013dxa,Eichmann:2017wil,Weil:2017knt} are required when working with the new pion kernels, as briefly described in Sec.~\ref{subsec:branchcut} and in Appendix \ref{sec:contour_deformation}. We then show our results for the quark-photon vertex in the complex plane and for purely spacelike photon momentum in Sec.~\ref{sec:Results}. We relegate the most technical aspects to several Appendices. Note that we work in Landau gauge and in Euclidean spacetime (using the conventions in \cite{Eichmann:2016yit}). 

\section{Formalism}
\label{sec:Formalism}

In this section we briefly summarise the elements of the DSE/BSE formalism necessary to study the quark-photon vertex. For further details we refer to \cite{Eichmann:2016yit,Sanchis-Alepuz:2017jjd}.

\begin{figure*}[ht]
\centerline{%
\includegraphics[width=8.5cm]{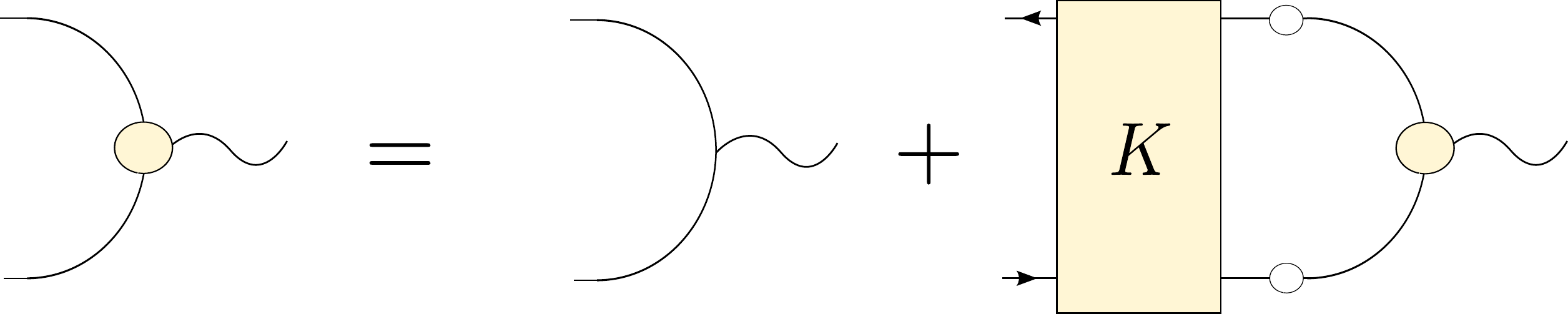}}
\caption{Inhomogeneous BSE for the quark-photon vertex. The first term on the right-hand side is the inhomogeneous term in Eq.~\eqref{eq:inhomBSE_vector} (also, $\Gamma_0$ in Eq.~\eqref{eq:formaldefVertex}), namely $Z_2\gamma^\mu t$. The BSE interaction kernel is denoted by K and lines with blobs represent full quark propagators.}
\label{Fig:inhomBSE}     
\end{figure*}

\subsection{Quark-photon vertex}
\label{sec:DSEsBSEs}

In the DSE/BSE formalism, the quark-photon vertex $\Gamma^\mu$ can be described by an integral equation which is linear in $\Gamma^\mu$, usually called inhomogeneous BSE (see Fig.~\ref{Fig:inhomBSE})\hfill
\begin{widetext}
\flaligne{
\bracket{\Gamma^{\mu}}_{a\alpha,b\beta}\bracket{Q,p}=&Z_2 \bracket{\gamma^\mu}_{ab} t_{\alpha\beta}+\int_q K^{r\rho,s\sigma}_{a\alpha,b\beta}\bracket{Q,p,q}S_{r\rho,e\epsilon}\bracket{k_1}\bracket{\Gamma^{i,\mu}}_{e\epsilon,n\nu}\bracket{Q,q}S_{n\nu,s\sigma}\bracket{k_2}~,\label{eq:inhomBSE_vector}
}
\end{widetext}
where $Q$ is the photon momentum, $p$ is the relative momentum between quark and antiquark, the internal relative momentum $q$ is integrated over and the internal quark and antiquark momenta are $k_1=q+Q/2$ and $k_2=q-Q/2$, respectively, so that $Q=k_1-k_2$ and $q=(k_1+k_2)/2$. The interaction between quark and antiquark is  described by the Bethe-Salpeter interaction kernel $K$ and $S$ is the full, non-perturbative quark propagator.
Latin letters represent Dirac indices and Greek letters represent flavour indices (we have  omitted colour indices for clarity) and $Z_2$ is the quark renormalisation constant (see below). The $SU(3)$ flavour structure of the vertex is given by $t_{\alpha\beta}=\textrm{diag}\bracket{\nicefrac{2}{3},\nicefrac{-1}{3},\nicefrac{-1}{3}}$.

Equation \eqref{eq:inhomBSE_vector} is obtained after projecting the quark-antiquark four-point Green's function $G$  onto the subspace with the quantum numbers of the photon. Namely, Dyson's equation for $G$ reads
\flaligne{G=G_0+G_0KG~,\label{eq:DysonG}}
with $G_0$ the product of a quark and an antiquark propagators and $K$, the interaction kernel defined above. The quark-photon vertex is defined as the projection
\flaligne{\Gamma^\mu\equiv G_0^{-1}G\Gamma_0^\mu~,\label{eq:formaldefVertex}}
where $\Gamma_0^\mu$ is any tensorial object with the appropriate quantum numbers; to derive \eqref{eq:inhomBSE_vector} we used for $\Gamma_0^\mu$ the tree-level vertex $\Gamma_0^\mu=Z_2\gamma^\mu t$.

Moreover, from the Green's function $G$ one can introduce the amputated scattering matrix $T$
\flaligne{G=G_0+G_0TG_0~,\label{eq:definitionT}}
and Eq.~\eqref{eq:DysonG} becomes
\flaligne{
T=K+KG_0T~.\label{eq:DysonT}
}
If the quark-antiquark system forms a resonance for a (possibly complex) rest-frame total momentum $P^2=-M^2+iM\Gamma$ the scattering matrix features then a pole
\flaligne{
 T\sim \frac{\Psi \bar{\Psi}}{P^2 + M^2 -iM\Gamma},
\label{eq:pole}
}
and expanding \eqref{eq:DysonT} around this pole and keeping the most singular terms only one obtains a homogeneous Bethe-Salpeter equation for $\Psi$
\flaligne{
\Psi=KG_0\Psi~.\label{eq:homogeneousBSE}
}
From \eqref{eq:definitionT} we easily get
\flaligne{\Gamma^\mu=\Gamma_0^\mu+TG_0\Gamma_0^\mu~,}
which clearly shows that, if the scattering matrix has bound-state poles for the quantum numbers of $\Gamma_0^\mu$, these will also manifest as poles in the quark-photon vertex. Note that the mass $M$ and width $\Gamma$ in \eqref{eq:pole} would be the result of solving a BSE and not some input of the calculation.

Note that in deriving \eqref{eq:inhomBSE_vector} from \eqref{eq:DysonG} we have implicitly assumed that the quark-photon vertex appears on the right hand side through $G$ only and not through the interaction kernel $K$. That is, \eqref{eq:inhomBSE_vector} is sufficient to describe the quark-photon vertex only in an approximation where one ignores the electromagnetic corrections to the vertex. Thereby, the interaction kernel in \eqref{eq:DysonG} contains only QCD interactions among quarks. 

In order to fully specify Eqs.~\eqref{eq:inhomBSE_vector} or \eqref{eq:homogeneousBSE} it is necessary to define the interaction kernel $K$ as well as the quark propagator $S(p)$. The latter is given as a solution of the quark DSE
\begin{equation}
S^{-1} = Z_2  S_0^{-1} - Z_{1f} \int_q \gamma_\mu S(q) \Gamma^{qgl}_\nu(q,k) D_{\mu \nu}(k)~,\label{eq:quarkDSE}
\end{equation}
with 
\begin{equation}
S_0^{-1}(p) = i \slashed{p} + m~,
\end{equation}
where $Z_{1f}$ and $Z_2$ are renormalisation constants, $m$ is the renormalisation-point dependent current quark mass, $D_{\mu \nu}$ is the full gluon propagator which reads (in Landau gauge)
\begin{equation}
D_{\mu \nu}(k) = \left( \delta_{\mu \nu} - \frac{k_\mu k_\nu}{k^2}\right) \frac{Z(k^2)}{k^2}~,
\end{equation}
with $Z(p^2)$ the gluon dressing function and, finally, $\Gamma^{qgl}$ is the full quark-gluon vertex. 

\begin{figure*}[ht]
\centerline{%
\includegraphics[width=14.5cm]{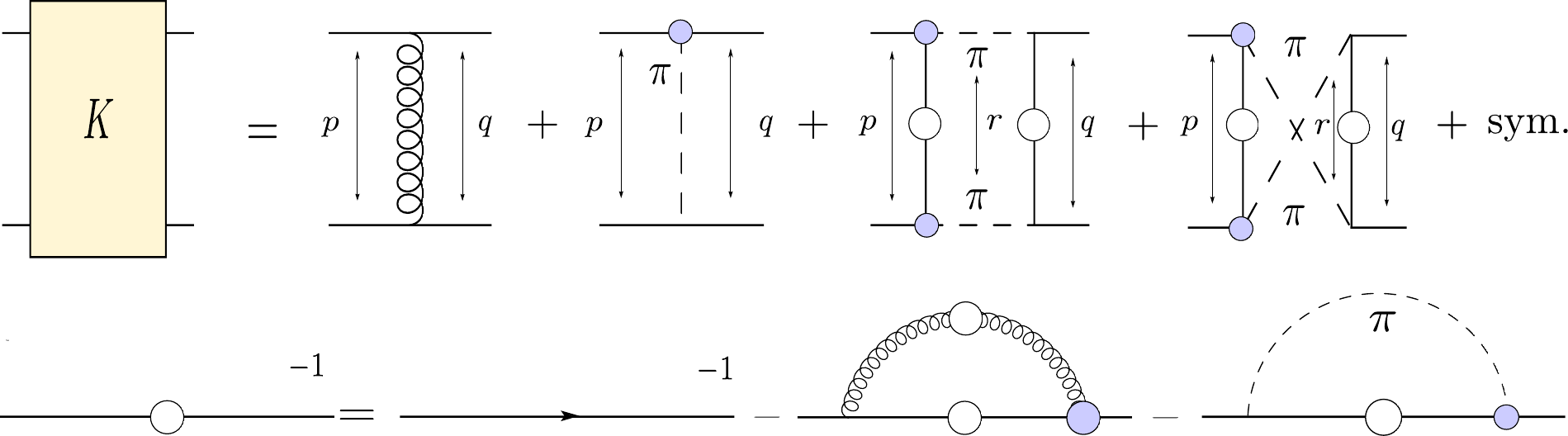}}
\caption{Interaction kernel including t, s and u pion exchange channels (upper), and the Dyson-Schwinger equations for the quark propagator with one gluon exchange and one pion exchange (lower), where $p,q$ and $r$  are the quark relative momentum. Vertices with and without blobs represent, respectively, full and bare interaction vertices. Similarly, lines with blobs represent full quark propagators.}
\label{Fig:newkernel}     
\end{figure*}

\subsection{Interaction kernels and symmetries}\label{sec:kernels}

The remaining element to be defined is the quark-antiquark interaction kernel $K$. A diagrammatic definition thereof consists of a sum of infinite terms, each describing one of the possible different interaction processes among quark and antiquark. This makes plain that, in practical calculations, the expansion of the interaction kernel must be truncated to a sum of a finite number of terms. These should be judiciously  chosen in order that the relevant physical effects are kept in the calculation.

The most widely used truncation of the BSE kernel $K$ is the so-called rainbow-ladder (RL) truncation. Here, the kernel consists of a vector-vector gluon exchange (see Fig.~\ref{Fig:newkernel}), namely (omitting again colour indices)
\flaligne{
K^{r\rho,s\sigma}_{a\alpha,b\beta}\bracket{Q,p,q}=\alpha\bracket{k^2}\gamma^\mu_{ar}\gamma^\nu_{sb}D^{\mu\nu}\bracket{k}\delta^{\alpha\rho}\delta^{\sigma\beta}~,
}
with $k=p-q$ the gluon momentum and $\alpha(k^2)$ an effective coupling that provides strength to the quark-antiquark interaction. To parametrise this effective interaction we use the following model \cite{Maris:1997tm, Maris:1999nt}
\begin{flalign}\label{eq:MTmodel}
\alpha(q^2) {}=&
 \pi\eta^7\left(\frac{q^2}{\Lambda^2}\right)^2
e^{-\eta^2\frac{q^2}{\Lambda^2}}\nonumber\\ &+{}\frac{2\pi\gamma_m
\big(1-e^{-q^2/\Lambda_{t}^2}\big)}{\textnormal{ln}[e^2-1+(1+q^2/\Lambda_{QCD}
^2)^2]}~.
\end{flalign}
The interaction strength is characterized by an energy
scale $\Lambda$ and a dimensionless parameter $\eta$ that controls the width of the interaction (the
scale $\Lambda_t=1$~GeV is introduced for technical reasons and has no impact on the results). These parameters will be fixed to reproduce correctly the pion
decay constant from truncated meson BSE. For the anomalous dimension we use $\gamma_m=12/(11N_C-2N_f)=12/25$,
corresponding to $N_f=4$ flavours and $N_c=3$ colours. For the QCD scale we take $\Lambda_{QCD}=0.234$ GeV.

In order to successfully apply the Bethe-Salpeter formalism to hadron phenomenology it is important that the relevant global symmetries are correctly implemented. In particular, chiral symmetry of the QCD Lagrangian with massless quarks ensures that pions are massless bound states in the chiral limit, as a consequence of Goldstone's theorem and dynamical chiral symmetry breaking. $U(1)$ vector symmetry of the QCD Lagrangian, on the other hand, ensures charge conservation, which is crucial for a correct calculation of hadron form factors. Whether these symmetries are preserved by BSEs when a truncated kernel is used is not guaranteed. Chiral symmetry will be correctly implemented only if the kernel fulfils the axial-vector Ward-Takahashi identity (Ax-WTI)
\flaligne{
&i\Sigma_{ar}\bracket{p_+}\gamma^5_{rb}t^i_{\alpha\beta}
+i\gamma^5_{ar}\Sigma_{rb}\bracket{p_-}t^i_{\alpha\beta}=\nonumber\\
&~~\int_q K^{r\rho,s\sigma}_{a\alpha,b\beta}\bracket{Q,p,q}\left[i~t^i_{\rho\nu}\gamma^5_{rn}S_{n\nu,s\sigma}\bracket{q_-} +\right.\nonumber\\
&~~~~~~~~~~\left.i~S_{r\rho,e\epsilon}\bracket{q_+}\gamma^5_{es}~t^i_{\epsilon\sigma}\right]~,\label{eq:AxWTI_BSEkernel}
}
with $\Sigma$ the quark self-energy (namely, the second term in the right-hand term of \eqref{eq:quarkDSE}) and $v_\pm=v\pm Q/2$. Similarly, vector symmetry will be correctly implemented if the kernel obeys the vector Ward-Takahashi identity (V-WTI) 
\flaligne{
&i\Sigma_{ab}\bracket{p_+}t^i_{\alpha\beta}
-i\Sigma_{ab}\bracket{p_-}t^i_{\alpha\beta}=\nonumber\\
&~~\int_q K^{r\rho,s\sigma}_{a\alpha,b\beta}\bracket{Q,p,q}\left[i~t^i_{\rho\nu}S_{r\nu,s\sigma}\bracket{q_-}-\right.\nonumber\\
&~~~~~~~~~~\left.i~S_{r\rho,s\epsilon}\bracket{q_+}~t^i_{\epsilon\sigma}\right]~.\label{eq:VWTI_BSEkernel}
}
Clearly now, a truncated kernel $K$ entails a certain approximation of the quark-gluon vertex, which appears implicitly via the quark self-energy $\Sigma$. In particular, the RL truncation of the BSE kernel respects the symmetries above if the quark-gluon vertex is simplified such that $Z_{1f}\gamma_\mu Z(k^2) \Gamma_\nu(q,k) \rightarrow Z_2^2\gamma_\mu \alpha(k^2) \gamma_\nu$ in the quark DSE \eqref{eq:quarkDSE}, with $\alpha(k^2)$ the effective interaction \eqref{eq:MTmodel}.

\subsection{The t-channel pion-exchange kernel}\label{subsec:tchannel}

As previously discussed, we are interested in studying the effect of non-valence quarks in the quark-photon vertex, as a first step towards the implementation of meson-cloud effects in the calculation of hadron form factors. It was argued in \cite{Fischer:2007ze,Fischer:2008sp,Sanchis-Alepuz:2014wea} that such \textit{unquenching} effects may be approximated by the inclusion of explicit pionic degrees of freedom in the interaction kernel $K$, in addition to quarks and gluons.

As a first term of this type consider the second diagram in the kernel depicted in Fig.~\ref{Fig:newkernel}, that is, the exchange of a pion between the quark and antiquark; we refer to this term in what follows as the t-channel pion exchange. Here the quark-pion vertex is taken to be the pion Bethe-Salpeter amplitude, which in full generality reads 
\flaligne{
\Gamma^i_{\pi}(p,P) &=\tau^i~\gamma_5 \{ E_{\pi}(p,P) - i \slashed{P} F_{\pi}(p,P) \nonumber \\
 &- i\slashed{p} (p \cdot P) G_{\pi}(p,P) - \left[ \slashed{P},\slashed{p}\right]H_{\pi}(p,P) \} ~\label{eq:pionamplitude}
}
with $P$ the pion momentum and $p$ the relative momentum between quark and antiquark, $\tau^i$ represents the flavour structure of the pion isovector and $E_{\pi}, F_{\pi} , G_{\pi}, H_{\pi}$ are four independent dressing functions. At the level of the quark DSE, the quark-pion interaction entails the addition of a pion loop, as depicted in Fig.~\ref{Fig:newkernel}.

It turns out, however, that the straightforward implementation of those pion-exchange terms does not respect the WTIs above. In Refs.~\cite{Fischer:2007ze,Fischer:2008sp} it was shown that the quark-pion interaction can be modified in a way such that the Ax-WTI is preserved. The modified t-channel BSE kernel then reads (with no flavour indices, see below)\hfill
\begin{flalign}
&K^{(t)~ut}_{rs}(q,p;P) =\nonumber\\
&~~~~~~~~~~~\frac{C}{4} [\Gamma_{\pi}^j]_{ru} \left(\frac{p + q - P}{2}; p - q \right) [Z_2 \gamma^5]_{ts} D_{\pi}(p - q) \nonumber \\ \nonumber
 &~~~~~~~~+\frac{C}{4} [\Gamma_{\pi}^j]_{ru} \left(\frac{p + q - P}{2}; q - p \right) [Z_2 \gamma^5]_{ts} D_{\pi}(p - q) \\ \nonumber
 &~~~~~~~~+\frac{C}{4} [Z_2 \gamma^5]_{ru} [\Gamma_{\pi}^j]_{ts} \left(\frac{p + q + P}{2}; p - q \right) D_{\pi}(p - q) \\ 
 &~~~~~~~~+\frac{C}{4} [Z_2 \gamma^5]_{ru} [\Gamma_{\pi}^j]_{ts} \left(\frac{p + q + P}{2}; q - p \right) D_{\pi}(p - q)~,\label{eq:BSEkernel_tchannel}
\end{flalign}
which preserves the Ax-WTI in combination with the following truncation of the quark DSE
\begin{eqnarray}
S^{-1}(p) &=& S^{-1}(p)^{RL} - \frac{3}{2} \int_q \Bigg[Z_2 \gamma_5 S(q) \Gamma_{\pi}\left(\frac{p+q}{2}, q-p\right) \nonumber \\
&&  + Z_2 \gamma_5S(q)\Gamma_{\pi}\left(\frac{p+q}{2}, p-q\right)\Bigg] \frac{D_{\pi}(k)}{2}~,\label{eq:quarkDSE_tchannel}
\end{eqnarray}
with $S^{-1}(p)^{RL}$ the quark DSE in the RL truncation previously described. In Eqs.~\eqref{eq:BSEkernel_tchannel} and \eqref{eq:quarkDSE_tchannel} the pion propagator is taken as $D_{\pi}(k) = (k^2 + m_{\pi}^2)^{-1}$. The factor $3/2$ in Eq.~\eqref{eq:quarkDSE_tchannel} stems from the flavour traces and so should the factor $C$ in \eqref{eq:BSEkernel_tchannel} be obtained. However, only if one imposes a value $C=-3/2$, in order to match the global factors of the quark DSE, is the Ax-WTI preserved. Unfortunately, such a choice for $C$, when done in the quark-photon vertex BSE for which flavour traces would lead to $C=+3/2$ instead, violates the V-WTI. Since it is the latter that we are most interested in preserving in the present investigation (since it relates to charge conservation in electromagnetic processes), we adhere herein to the interpretation of $C$ as a flavour factor, at the expense of violating the Ax-WTI. We find this acceptable since the goal of the present investigation is a qualitative understanding of the analytic structure of the vertex and the changes thereof induced by the kernels below. 

\subsection{The s- and u-channel pion exchange}\label{subsec:suchannel}

The s- and u-channels (resp. third and fourth terms in Fig.~\ref{Fig:newkernel}) were also introduced in \cite{Fischer:2007ze}, as they are of the same order as the t-channel contribution above in a resummed $N_c$-expansion of the effective action. Their effect on BSE calculations was, however, not considered in subsequent calculations. This was recently studied in \cite{Williams:2018adr}, where it was shown that adding those terms to the BSE kernel enables virtual decay channels in the rho-meson channel and, thus, obtaining for the first time a bound-state with a non-vanishing width as a solution of a BSE calculation. 

We used here a slightly different definition of the kernels in the s- and u-channels in order to be consistent with the construction for the t-channel, where one of the pion kernels is kept bare and the kernel is then symmetrised. They thus read\hfill
\begin{widetext}
\flaligne{
K^{(s)~he}_{da}(q,p,r;P)=~& \frac{C}{2}~D_{\pi}\left(\frac{P + r}{2}\right) D_{\pi}\left(\frac{P - r}{2}\right)~\left[[Z_2\gamma_5]_{dc} S_{cb}\left(p - \frac{r}{2}\right)[Z_2\gamma_5]_{ba}\right.\nonumber\\
&~~~~\times[\Gamma_{\pi}^j]_{hg} \left(q - \frac{P}{4} - \frac{r}{4}; \frac{r - P}{2} \right) S_{gf}\left(q - \frac{r}{2}\right)[\Gamma_{\pi}^j]_{fe} \left(q + \frac{P}{4} - \frac{r}{4}; -\frac{P + r}{2} \right)\nonumber\\
~&+~[\Gamma_{\pi}^j]_{dc} \left(p + \frac{P}{4} - \frac{r}{4}; \frac{P + r}{2} \right) S_{cb}\left(p - \frac{r}{2}\right)[\Gamma_{\pi}^j]_{ba} \left(p - \frac{P}{4} - \frac{r}{4}; \frac{P - r}{2} \right)\nonumber\\
&~~~~\left.\times[Z_2\gamma_5]_{hg} S_{gf}\left(q - \frac{r}{2}\right)[Z_2\gamma_5]_{fe}\right]~, 
\label{eq:BSEkernel_schannel}
\\
K^{(u)~he}_{da}(q,p,r;P)=~& \frac{C}{2}~D_{\pi}\left(\frac{P + r}{2}\right) D_{\pi}\left(\frac{P - r}{2}\right)~\left[ [Z_2\gamma_5]_{dc}  S_{cb}\left(p + \frac{r}{2}\right) [Z_2\gamma_5]_{ba}\right.   \nonumber \\
 &~~~~\times [\Gamma_{\pi}^j]_{hg} \left(q - \frac{P}{4} - \frac{r}{4}; \frac{r - P}{2} \right) S_{gf}\left(q - \frac{r}{2}\right) [\Gamma_{\pi}^j]_{fe} \left(q + \frac{P}{4} - \frac{r}{4}; -\frac{P + r}{2} \right)\nonumber\\
~&+~[\Gamma_{\pi}^j]_{dc} \left(p + \frac{P}{4} + \frac{r}{4}; \frac{P - r}{2} \right) S_{cb}\left(p + \frac{r}{2}\right)[\Gamma_{\pi}^j]_{ba} \left(p - \frac{P}{4} + \frac{r}{4}; \frac{P + r}{2} \right)  \nonumber \\
 &~~~~\left.\times[Z_2\gamma_5]_{hg} S_{gf}\left(q - \frac{r}{2}\right)[Z_2\gamma_5]_{fe}\right]~,
\label{eq:BSEkernel_uchannel}
}
\end{widetext}
where now $r$ is an additional integration momentum in the BSE (cf. Eqs.\eqref{eq:inhomBSE_vector} or \eqref{eq:homogeneousBSE}). 
These kernels, when considered as stemming from a resummed effective action, do not lead to additional contributions in the quark DSE (see \cite{Fischer:2007ze}). Note also that they do not contribute to the Ax-WTI, as can be easily checked with symmetry considerations. On the other hand, from flavour traces we obtain $C=+3/2$ for these kernels as well, which suffices to preserve the V-WTI. 

\subsection{Branch-cut structure}\label{subsec:branchcut}

\begin{figure*}[htb!]
\centerline{%
\includegraphics[width=14.5cm]{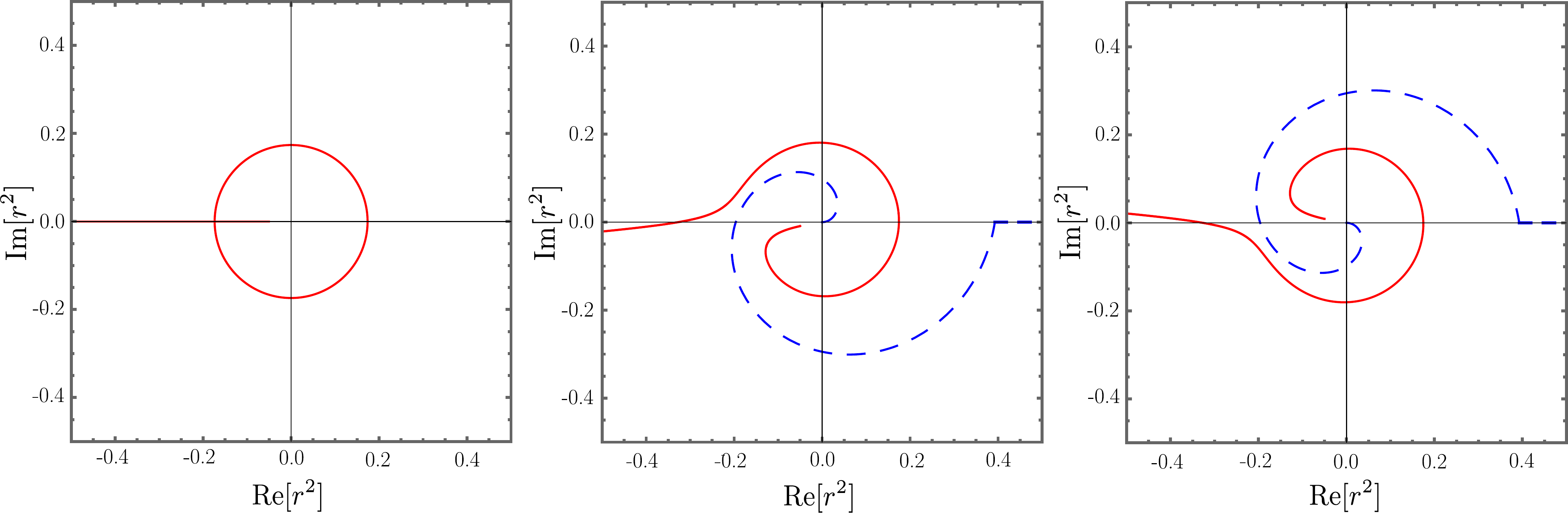}}
\caption{Branch cuts that appear in the s- and u-channel pion kernels after angular integration. The solid line corresponds to the branch cuts due to the two pion propagators. For purely real $Q^2$ at the two-pion production threshold $-4m_\pi^2$ and below (left), the branch cut is closed and overlaps the real and positive $r^2$ half-axis. We add a positive (middle) or negative (right) imaginary part to $Q^2$ to get an opening in the cuts. The long-dashed line shows a possible $r^2$ integration path. }
\label{Fig:cuts}   
\end{figure*}

\begin{figure*}[htb!]
\centerline{%
\includegraphics[width=10cm]{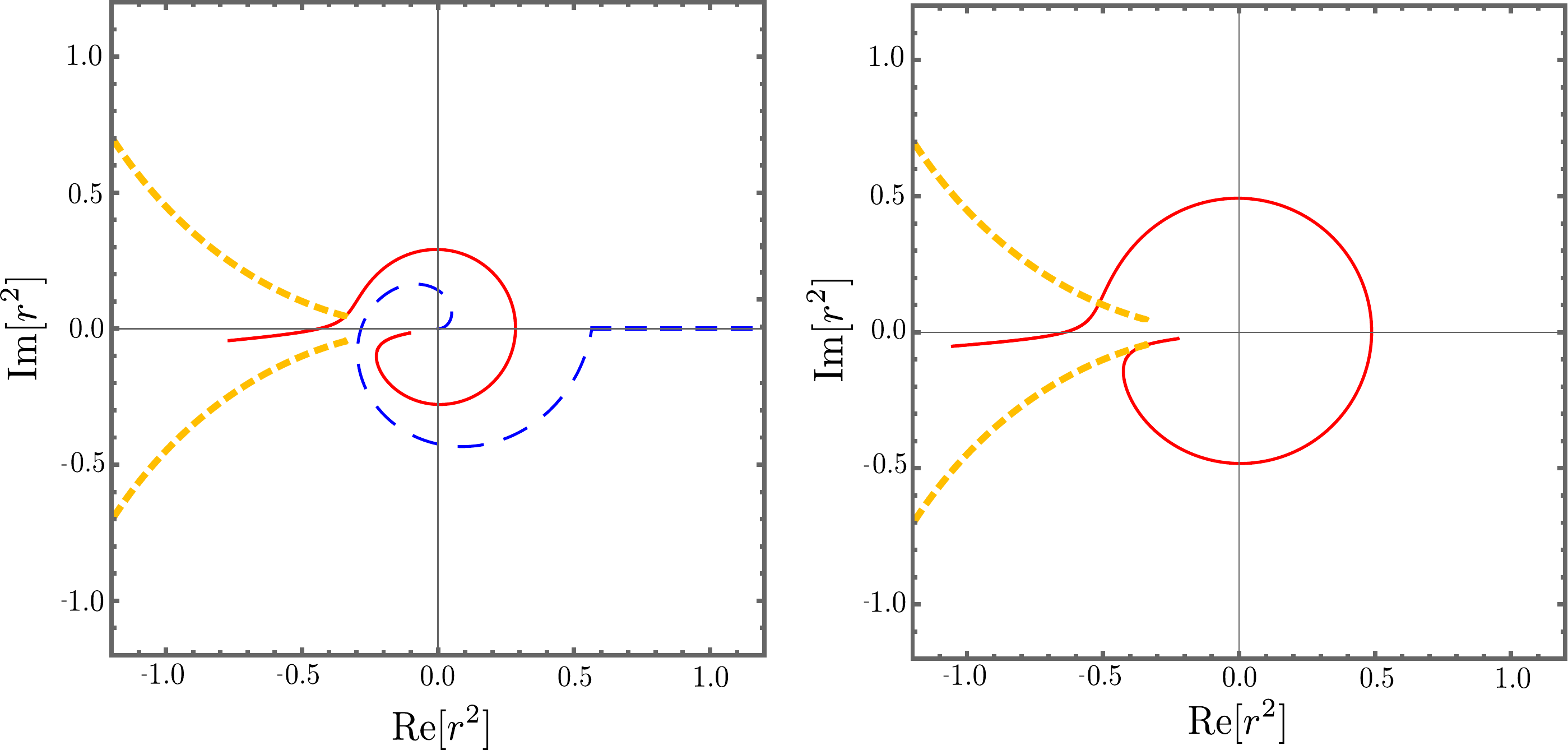}}
\caption{Branch cuts in the s- and u-channel pion kernels stemming from the pion propagators (solid lines) and from the quark propagators (dashed lines), for two different values of the photon momentum $Q^2=Q^2_r+iQ^2_i$. The left panel corresponds to $Q^2_r=-0.25$~GeV$^2$ and $Q^2_i=0.025$~GeV$^2$ and the right panel corresponds to $Q^2_r=-0.49$~GeV$^2$ and $Q^2_i=0.035$~GeV$^2$. The long-dashed line in the left panel shows a possible $r^2$ integration path. No integration path exists for the case in the second panel, since there the cuts from the quark and pion propagators overlap.}
\label{Fig:cutspoles}   
\end{figure*}

The inclusion of the two kernels given in equations (\ref{eq:BSEkernel_schannel}) and (\ref{eq:BSEkernel_uchannel}) poses an additional challenge in BSE calculations. The reason is their non-trivial analytic structure, induced by the intermediate pions going potentially on-shell as well as by singularities in the quark propagators. For example, the kernel features now branch cuts, generated by the pion propagators upon $r$-integration. The position of those cuts can be determined by studying the zeroes of the denominators
\begin{eqnarray}
y_1(P,z_r) &=& -m_{\pi}^2 - P^2 + 2 z_r^2 P^2 \nonumber\\
 &-& 2\sqrt{-m_{\pi}^2 z_r^2 P^2 - z_r^2 P^4 +  z_r^4 P^4} \nonumber \\
y_2(P,z_r) &=& -m_{\pi}^2 - P^2 + 2 z_r^2 P^2 \nonumber \\
&+& 2\sqrt{-m_{\pi}^2 z_r^2 P^2 - z_r^2 P^4 +  z_r^4 P^4} 
\end{eqnarray}
where we assumed a standard expression of $r$ in hyperspherical coordinates
\flaligne{r=\sqrt{r^2}&\left(\sqrt{1-z_r^2}~\sqrt{1-y_r^2}~\sin\phi_r,\right.\nonumber \\
                      &\left.\sqrt{1-z_r^2}\sqrt{1-y_r^2}~\cos\phi_r,\sqrt{1-z_r^2}~y_r,z_r\right)~,}
with $r^2\in[0,\infty)$, $y_r,z_r\in[-1,1]$ are cosines of angles and $\phi_r\in[0,2\pi)$. A branch cut in the complex $r^2$-plane appears upon integration of the $z_r$ coordinate, as can be seen in Fig.~\ref{Fig:cuts} for different values of the total momentum the BSE (eg. $Q$ in Eq.~\eqref{eq:inhomBSE_vector}). For $Q^2>-4m_\pi^2$ the cut is entirely in the negative-$r^2$ half-plane and, thus, does not interfere with the $r^2$-integration. Below the two-pion production threshold (i.e. $Q^2\le -4m_\pi^2$), however, the cut moves into the positive-$r^2$ half-plane, obstructing the $r^2$-integration. In order to be able to perform the $r^2$-integration in this case, one must deform the integration path \cite{Windisch:2012sz,Windisch:2013dxa,Eichmann:2017wil,Weil:2017knt}, as sketched in Fig.~\ref{Fig:cuts} (for details see Appendix \ref{sec:contour_deformation}). This is only possible, however, if the branch cut is open which only happens if $Q^2$ has an imaginary part that is interpreted, in the context of homogeneous BSEs, as the decay width of the state (see Eq.~\eqref{eq:pole}). When solving the inhomogeneous BSE for the quark-photon vertex, the choice of the imaginary part of $Q^2$ selects different Riemann sheets in the $Q^2$ plane, as we will see below. 

Note also that in both truncations of the quark DSE used herein, the quark propagator features pairs of complex conjugate poles in the left complex half-plane. These induce additional branch cuts as depicted in Fig.~\ref{Fig:cutspoles}. Those cuts limit the maximum value of $abs(-Q^2)$ that we can access in our calculations.

\section{Results}\label{sec:Results}

We show here the results for the quark-photon vertex using the three different truncations of the BSE kernel discussed so far, starting with the RL truncation and then adding the t-channel pion exchange and finally the s- and u-channel terms. The parameters of the effective interaction \eqref{eq:MTmodel} are fixed to be $\eta=1.44$ and $\Lambda=0.74$, which leads to a pion mass $m_\pi=138$~MeV and a pion decay constant $f_\pi=129$~MeV as a solution of the homogeneous BSE with a kernel consisting of a RL term and a t-channel pion exchange kernel (since the s- and u-channel terms do not contribute to the pion BSE) and for a quark mass $m_q=6.3$~MeV at a renormalisation scale $\mu=19$~GeV. Incidentally, these parameters give a rho mass $m_\rho=736.5$~MeV and a decay constant $f_\rho=214$~MeV for the same truncation of the BSE kernel.

In what follows we show the dressing functions corresponding to an expansion of the full quark-photon vertex in a tensorial basis. It is advantageous to distinguish the purely transverse (with respect to the total momentum $Q$) from the rest, since only in the former will vector-meson bound-state poles appear due to the Proca condition (see discussion below)
\begin{align}
\Gamma^\mu(Q,p) =& \Bigl[\lambda_1(Q,p)\gamma^\mu + 2 p^\mu\left(\lambda_2(Q,p) \slashed{p} + i\lambda_3(Q,p)\right)\Bigr] +\nonumber\\ &\Gamma^\mu_T(Q,p)~,\label{eq:vertex_expansion1}
\end{align}
where $\lambda_i$ are scalar dressing functions that depend on momentum scalars $p^2$, $Q^2$ and $\hat{p}\cdot\hat{Q}$ and $\Gamma^\mu_T$ is the purely transverse ($Q_\mu\Gamma^\mu_T=0$) part of the vertex, which can in turn be expanded in an eight-dimensional covariant basis $\Gamma^\mu_T=\sum h_i\tau^i$, with $h_i$ scalar dressing functions. For illustration purposes we use the basis \cite{Eichmann:2014qva} (see Appendix \ref{sec:appendix_basis})
\begin{flalign}
\tau_1^\mu &=t^{\mu\nu}_{QQ}\gamma^\nu\;,&\tau_2^\mu &= t^{\mu\nu}_{QQ} \left(p\cdot Q\right) \frac{\mathrm{i}}{2}\left[\gamma^\nu,\slashed{p}\right]\;,\nonumber\\
\tau_3^\mu &=\frac{\mathrm{i}}{2}\left[\gamma^\mu,\slashed{Q}\right]\;,&\tau_4^\mu &=\frac{1}{6}\left[\gamma^\mu,\slashed{p},\slashed{Q}\right]\;,\nonumber\\
\tau_5^\mu &=t^{\mu\nu}_{QQ} \mathrm{i}p^\nu\;,&\tau_6^\mu &=t^{\mu\nu}_{QQ}p^\nu\slashed{p}\;,\nonumber\\
\tau_7^\mu &=t^{\mu\nu}_{Qp}\left(p\cdot Q\right)\gamma^\nu\;,&\tau_8^\mu &=t^{\mu\nu}_{Qp}\frac{\mathrm{i}}{2}\left[\gamma^\nu,\slashed{p}\right]\label{eq:vertex_expansion2}\;.
\end{flalign}
where $t^{\mu\nu}_{ab}=\left(a\cdot b\right) \delta^{\mu\nu}-b^\mu a^\nu$ and the triple commutator is defined as $[A,B,C]=[A,B]C + [B,C]A + [C,A]B$.

As a consequence of the V-WTI, the non-transverse dressing functions are uniquely given by the quark propagator dressing functions via the Ball-Chiu construction \cite{Ball:1980ay}. Comparing our numerical solutions for $\lambda_i$ with the Ball-Chiu vertex we have a consistency check that our truncation indeed preserves the V-WTI. 

Finally, to simplify the calculations whilst still keeping the physical features we wish to elucidate, namely the effect of intermediate on-shell particles in the BSE kernel, for the pion vertices appearing in the kernels we have only considered the leading $\gamma_5$ component of Eq.~\eqref{eq:pionamplitude}.

\subsection{Timelike photon momentum}

\begin{figure*}[htb!]
\centerline{%
\includegraphics[width=0.8\textwidth]{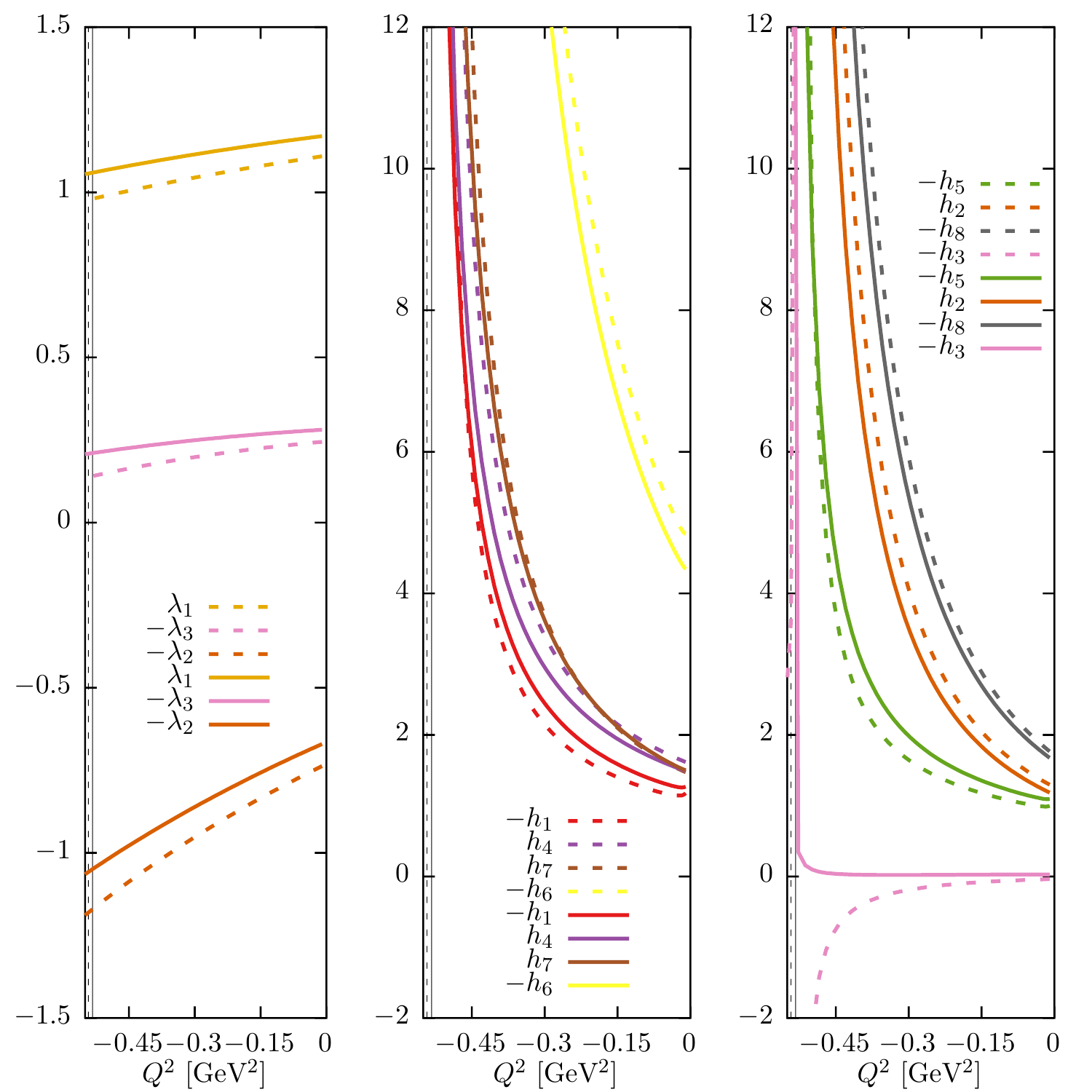}}
\caption{Dressing functions for the non-transverse ($\lambda_i$) and transverse ($h_i$) components (see Eqs.~\eqref{eq:vertex_expansion1} and \eqref{eq:vertex_expansion2}) of the quark-photon vertex in the $Q^2<0$ region for the RL truncation (solid lines) and with the addition of a t-channel pion exchange (dashed lines). The vertical solid and dashed lines indicate the position of the rho mass ($-m_\rho^2$), as obtained from the solution of an homogeneous BSE with the same truncated kernels.}
\label{Fig:qpv_Mink}     
\end{figure*}

We begin by studying the quark-photon vertex in the timelike domain. More precisely, we calculate it in a region of the complex $Q^2$-plane with $Re\bracket{Q^2}< 0$. As discussed in Sec.~\ref{subsec:branchcut}, the accessible region is limited by the position of the first pair of complex-conjugate poles of the quark propagator.

In this region, the vertex is sensitive to the quark-antiquark bound states with the quantum numbers of the photon $J^{PC}=1^{--}$, which include the rho meson and its excitations. This is manifested by the appearance of poles of the vertex dressing functions for the values of $Q^2$ for which the homogeneous BSE has solutions. In general, those solutions should be resonances and the pole occurs for complex values of $Q^2$ corresponding to the pole mass $Q^2=-M^2+iM\Gamma$, with $M$ and $\Gamma$ the Breit-Wigner mass and width of the resonance, respectively. That is, the analytic structure of the quark-photon vertex should feature isolated poles. Additionally, the possible decay modes in a given Green's function manifest themselves as the typical multiparticle branch cut, starting at the particle production threshold. 

The general features discussed above may or may not be present in the actual analytic structure of the solutions of a truncated BSE, depending on whether the BSE interaction kernel $K$ allows any decay mechanisms for the bound state (and thus there is at least the possibility of the BSE describing a bound state as a decaying resonance). 

In Fig.~\ref{Fig:qpv_Mink} we show the dressing functions corresponding to the longitudinal (non-transverse) and transverse components of the quark-photon vertex (cf. Eqs.~\eqref{eq:vertex_expansion1} and \eqref{eq:vertex_expansion1}) for a RL kernel as well as with the effects of a pion exchange (t-channel) included. Neither of those truncations of the BSE kernel allows a quark-antiquark bound state to decay, in the sense that the kernel does not contain any intermediate particles that may, potentially, go on-shell. This implies that the homogeneous BSE has bound-state solutions for real and negative values of the total momentum squared $P^2$ and, correspondingly, the dressing functions of the quark-photon vertex have bound-state poles for real and negative values of $Q^2$ only.

This is clearly seen in Fig.~\ref{Fig:qpv_Mink}. As indicated above, the bound states in the $1^{--}$ channel are the vector mesons which, due to the Proca condition, overlap only with the transverse components of the quark-photon vertex. Their dressings, therefore, feature poles for $Q^2=-M_\rho^2$, with $M_\rho$ the mass of the rho meson in the corresponding truncated homogeneous BSE. The dressings of the non-transverse components are, accordingly, regular for those $Q^2$ values. In fact, as discussed above, if the truncation of the DSE/BSE system preserves the vector WTI, then the non-transverse part of the quark-photon vertex is given by the Ball-Chiu vertex and, thus, can only feature the non-analyticities of the quark propagator.

\begin{figure*}[htb!]
\centerline{%
\includegraphics[width=0.8\textwidth]{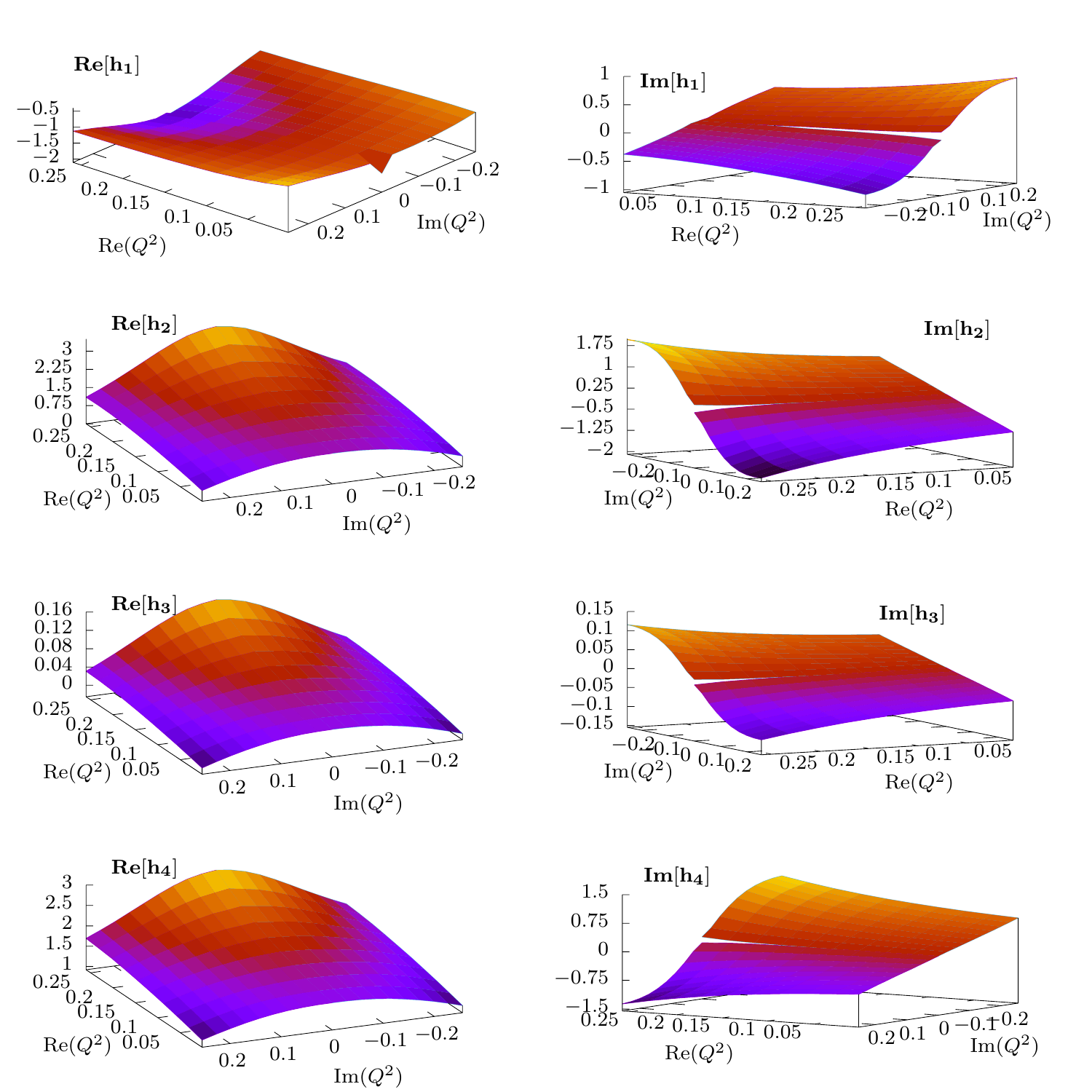}}
\caption{Real and imaginary parts of the transverse dressing functions $h_1$-$h_4$ of the quark-photon vertex, for $\hat{p}\cdot\hat{Q}=0$, in a region of the complex $Q^2$-plane with  $Re(Q^2)<0$ for the full truncation depicted in Fig~\ref{Fig:newkernel}. Even though not clearly visible in the plots, the branch cut in the imaginary parts begins at $Q^2=-4m_\pi^2$.}
\label{Fig:qpv_complex_trans14}     
\end{figure*}
\begin{figure*}[htb!]
\centerline{%
\includegraphics[width=0.8\textwidth]{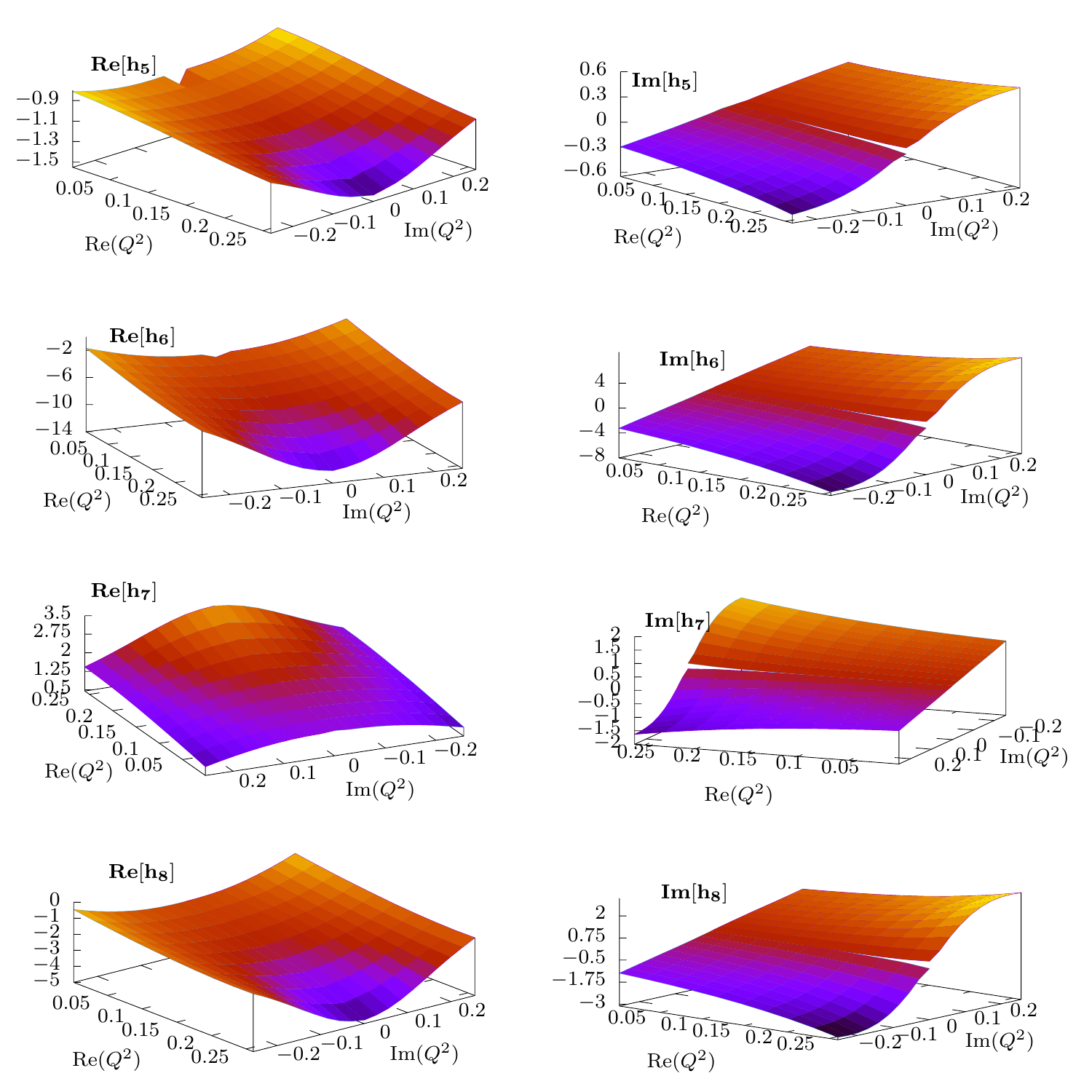}}
\caption{Real and imaginary parts of the transverse dressing functions $h_5$-$h_8$ of the quark-photon vertex, for $\hat{p}\cdot\hat{Q}=0$, in a region of the complex $Q^2$-plane with  $Re(Q^2)<0$ for the full truncation depicted in Fig~\ref{Fig:newkernel}. Even though not clearly visible in the plots, the branch cut in the imaginary parts begins at $Q^2=-4m_\pi^2$.}
\label{Fig:qpv_complex_trans58}     
\end{figure*}

We turn now to the discussion of the effect of the last diagrams (s- and u-channel) in Fig.~\ref{Fig:newkernel} on the analytic structure of the quark-photon vertex. In this case, the two intermediate pions go on-shell when $Q^2=-4m_\pi^2$ and thus represent, in particular, the $\rho\rightarrow\pi\pi$ decay channel. From the general discussion above, the dressings of the transverse components should feature a branch cut starting at the real and negative branch point $Q^2=-4m_\pi^2$. This is seen in Figs.~\ref{Fig:qpv_complex_trans14} and \ref{Fig:qpv_complex_trans58} as a discontinuity in the imaginary part of the dressing functions. Though not visible in the plots, the branch cut indeed opens at the expected value of $Q^2=-4m_\pi^2$ since, starting at this point and as discussed in Sec.~\ref{subsec:branchcut} and appendix \ref{sec:contour_deformation}, it is necessary to deform the BSE integration contour in order to avoid the non-analyticities of the BSE kernel induced by the presence of on-shell pions.  

As shown in \cite{Williams:2018adr}, the full BSE kernel studied herein is capable of partially describing the resonance character of the rho meson as a solution of the homogeneous BSE. In the present context, this simply means that the rho-meson bound-state pole of the transverse dressings of the quark-photon vertex, appearing for real $Q^2$ values in Fig.~\ref{Fig:qpv_Mink}, moves to the second Riemann sheet of the Riemann surface that is now the domain of the dressing functions. Unfortunately, with our choice of $\eta$ and $\Lambda$ parameters of the effective interaction \eqref{eq:MTmodel}, this pole is beyond the region we can access numerically. Its presence, however, can be inferred from the rise of the real part of the transverse dressing functions, see Figs.~\ref{Fig:qpv_complex_trans14} and \ref{Fig:qpv_complex_trans58}.

\begin{figure*}[htb!]
\centerline{%
\includegraphics[width=0.8\textwidth]{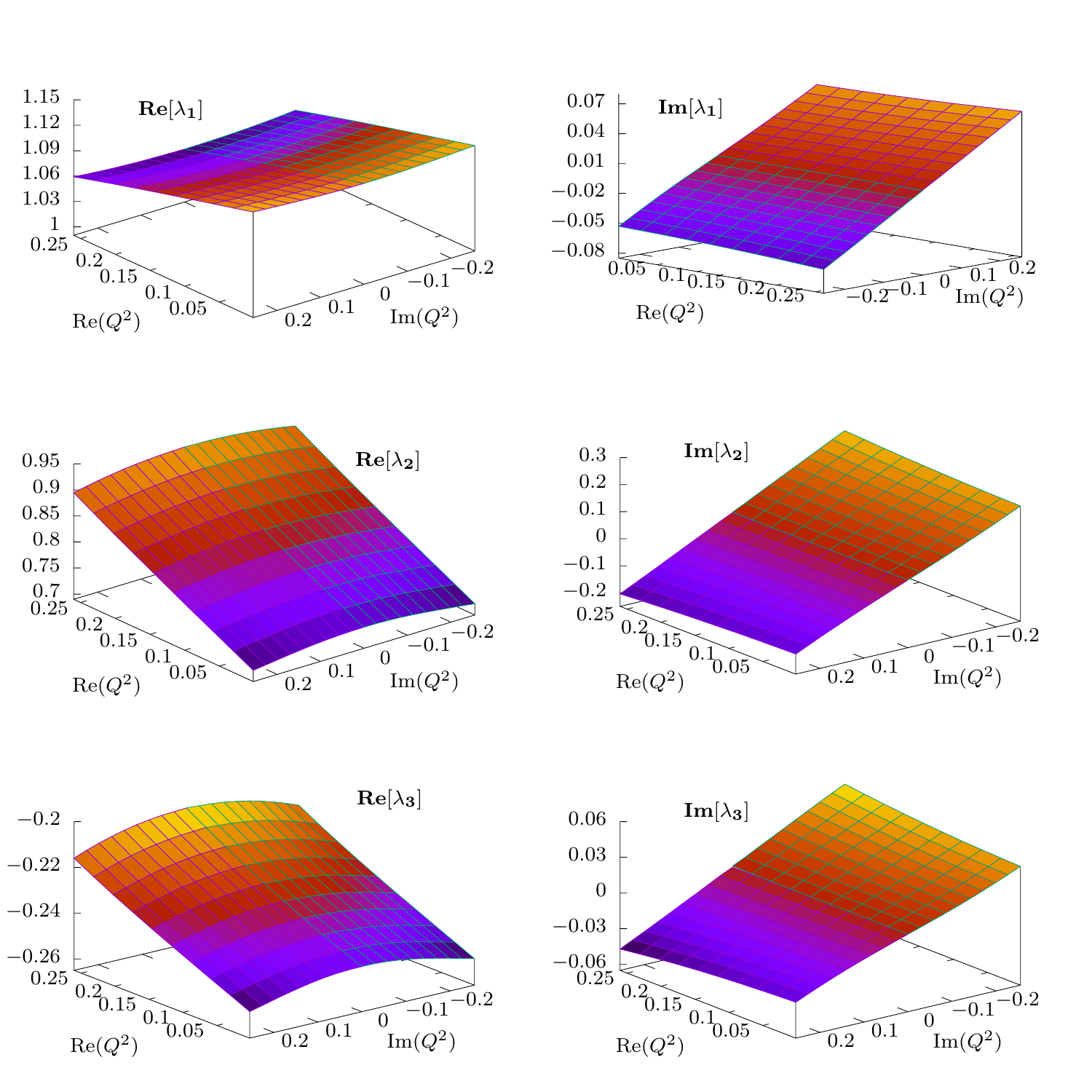}}
\caption{Real and imaginary parts of the non-transverse dressing functions $\lambda_1$-$\lambda_3$ of the quark-photon vertex, for $\hat{p}\cdot\hat{Q}=0$, in a region of the complex $Q^2$-plane with  $Re(Q^2)<0$ for the full truncation depicted in Fig~\ref{Fig:newkernel}.}
\label{Fig:qpv_complex_long}     
\end{figure*}

In Fig.~\ref{Fig:qpv_complex_long} we show our results for the non-transverse dressing functions. Since we adjusted the BSE kernel such that it preserves the vector WTI, the non-transverse part of the kernel is also given in this case by the Ball-Chiu vertex. Therefore, the corresponding dressing functions show a featureless behaviour in the complex $Q^2$ region studied in this work. We wish to stress again, that for all kernels discussed we have solved the equations for the twelve dressing functions describing the quark-photon vertex and we have only compared them with the Ball-Chiu expression as a check of our numerical calculations.

Finally, let us mention that the irregular behaviour of some dressing functions (e.g. $h_1$ in Fig.~\ref{Fig:qpv_complex_trans14}) in the vicinity of $Q^2=0$ is simply a numerical artefact stemming from the rotation of the basis we use for the calculation (see Appendix \ref{sec:appendix_basis}) onto the basis \eqref{eq:vertex_expansion2} used in the figures.

\subsection{Spacelike photon momentum}

\begin{figure*}[htb!]
\centerline{%
\includegraphics[width=0.8\textwidth]{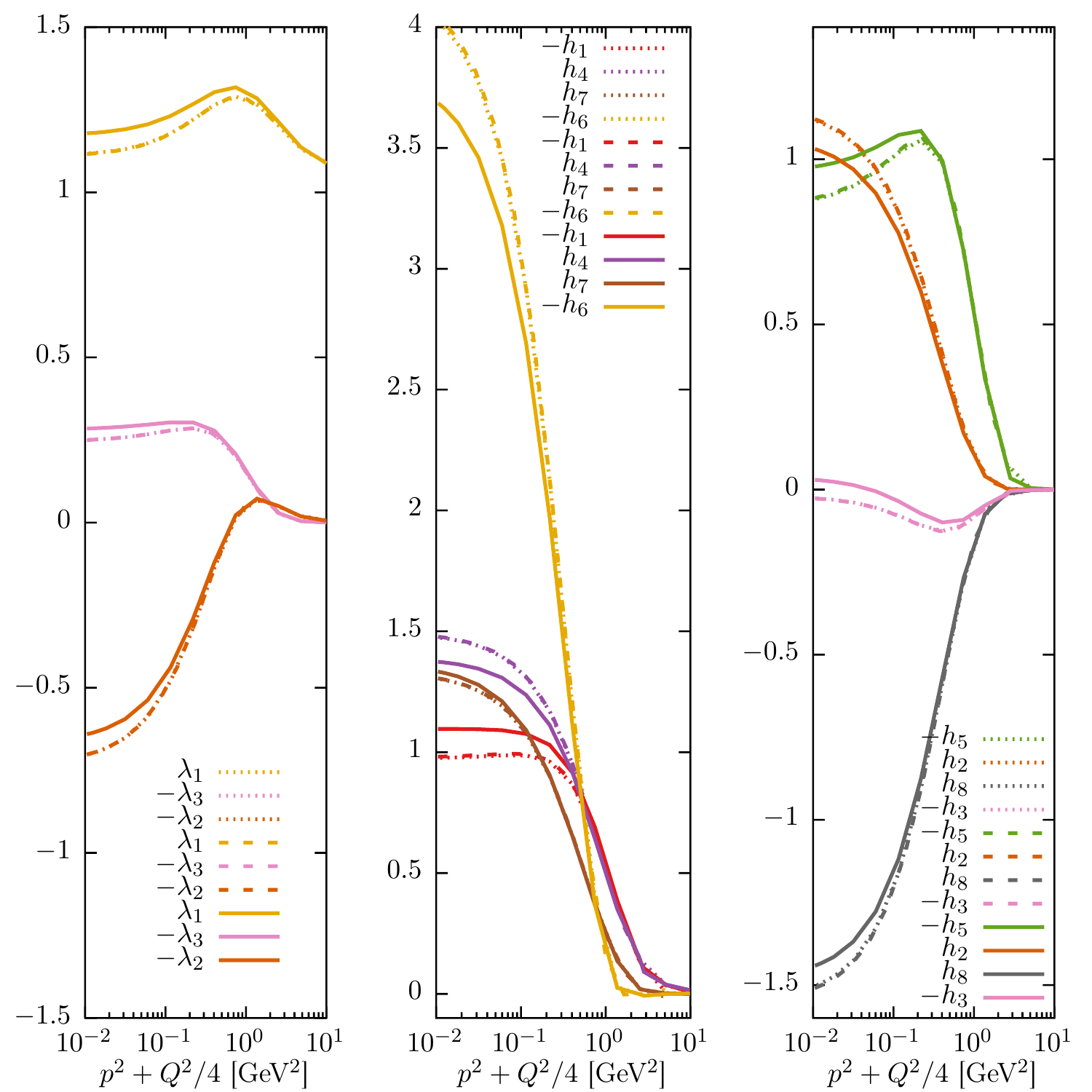}}
\caption{Dressing functions for the non-transverse ($\lambda_i$) and transverse ($h_i$) components (see Eqs.~\eqref{eq:vertex_expansion1} and \eqref{eq:vertex_expansion2}) of the quark-photon vertex in the spacelike $Q^2>0$ region for the RL truncation (solid lines), with the addition of a t-channel pion exchange (dashed lines) and for the full kernel in Fig.~\ref{Fig:newkernel} (dotted lines).}
\label{Fig:qpv_spacelike}     
\end{figure*}

For the calculation of spacelike hadron form factors (see, e.g. \cite{Eichmann:2016yit}), the quark-photon vertex is needed in the region $Q^2>0$. For this momentum region, first lattice QCD results have been published \cite{Tang:2019zbk} and compared with the results of a RL-truncated BSE calculation \cite{Eichmann:2014qva,Sanchis-Alepuz:2017jjd}. It is remarkable that the RL truncation shows already qualitative agreement with lattice QCD, a fact which certainly lies behind the success of spacelike form factor calculations in the BSE approach. Quantitatively, lattice QCD results show a weaker enhancement of the dressing functions at low $Q^2$, as compared to the RL results of \cite{Eichmann:2014qva,Sanchis-Alepuz:2017jjd}.

In Fig.~\ref{Fig:qpv_spacelike} we show our results for the transverse and non-transverse dressing functions for the RL truncation of the BSE kernel (solid lines), with the addition of a t-channel pion exchange (dashed lines) and for the full kernel which includes s- and u-channel pion terms as well. First of all, note that the difference between the RL results herein and those presented in \cite{Eichmann:2014qva,Sanchis-Alepuz:2017jjd} is solely due to the different choice of the parameters $\eta$ and $\Lambda$ for the effective interaction \eqref{eq:MTmodel}. Interestingly, our choice of parameters decreases the enhancement of some of the dressings for low photon momentum, as lattice QCD results indicate.

Upon inclusion of a pion-exchange in the t-channel, the dressings are slightly modified but none of them shows any qualitative change. More interesting is the situation when s- and u-channel pionic contributions are considered. We have seen in the previous subsection that the analytic structure of the quark-gluon vertex changes dramatically when these terms are included in the BSE kernel, from having poles on the negative real axis to featuring at least a multiparticle branch cut plus additional resonance poles in the complex plane. However, this barely reflects on the spacelike side, where the difference between dashed (t-channel) and dotted (s- and u-channel) curves is negligible. This is again encouraging and in line with our comment above about the suitability of the RL truncation for the calculation of hadron form factors; the qualitative features of the quark-photon vertex on the spacelike region appear to be rather insensitive to improvements on the RL kernel and, in particular, to qualitative changes in the timelike side. 

\section{Summary}

In this work we have studied the non-perturbative structure of the quark-photon interaction vertex in the spacelike photon momentum region $Q^2>0$ and in a region of the complex $Q^2$-plane with $Re(Q^2)<0$, for three different truncations of the inhomogeneous BSE that describes it. To the simple RL truncation we have added quark-pion interactions both as a t-channel pion exchange among quark and antiquark as well as s- and u-channel pion-emission channels. By using explicit pionic degress of freedom in addition to quarks and gluons, we aim at partially describing \textit{unquenching} effects on the quark DSE and the BSE kernel such as, of particular interest for the present study, the inclusion of decay channels in BSE calculations (these, without using pion degrees of freedom would be generated by the creation of a quark-antiquark pair inside the kernel, which could form on-shell bound states with the valence quarks). We have also discussed that the pionic kernels used herein are not entirely satisfactory since, in particular, they cannot be made preserve all the relevant global symmetries of QCD. However, they suffice to study the physical mechanism of interest for our investigation, namely the effect on the quark-photon vertex of intermediate on-shell particles in a BSE kernel.

We have seen how the effect of those intermediate particles is to drastically change the analytic structure of the quark-photon vertex. Whilst for the RL truncation, as well as with the inclusion of a t-channel pion exchange, the vertex only has poles for real and negative values of the photon momentum $Q^2$, upon inclusion of s- and u-channel pion kernels a multiparticle branch cut starting at $Q^2=-4m_\pi^2$ appears and the poles move away of the real axis into the complex plane. This is a reflection of the fact that for RL kernels (or similar) bound-state solutions of the homogeneous BSE are stable and hence they correspond to poles of suitable Green's functions for real and negative total momentum squared, whilst if a decay channel is available in the BSE kernel then the pole occurs for total momenta with an imaginary part, which corresponds to the decay width of the state. Our results for the analytic structure of the vertex are certainly closer to the expected physical picture and, hence, constitute a step forward towards the calculation of timelike hadron form factors in the BSE formalism.

On the spacelike $Q^2$ region this new analytic structure does not have any significant impact on the quark-photon vertex. This is reassuring in the sense that, even though an RL kernel does not generate the correct physical picture necessary for timelike form factor calculations, on the spacelike side this is not relevant and one can thus rely on the previous calculations of spacelike form factors using the RL truncation. 

\section{Acknowledgments}
This work was supported by the project P29216-N36 and the Doctoral Program W1203-N16 “Hadrons in Vacuum, Nuclei and
Stars”, both from the Austrian Science Fund, FWF. We thank R. Alkofer and A. Maas for fruitful discussions and for a critical reading of the manuscript.

\appendix

\section{Deformation of the integration path}\label{sec:contour_deformation}

To solve the inhomogeneous BSE when we include the s- and u-channel kernels in Eqs.~\eqref{eq:BSEkernel_schannel} and \eqref{eq:BSEkernel_uchannel} we need to perform an additional integration over the $r$-variable. For example, for a homogeneous BSE we must solve 
\begin{eqnarray}
\Psi(p;Q) &=& \int \frac{r^2  dr^2}{(2\pi)^4} \int \sqrt{1 -z_r}  dz_r  \int d\phi_r \int dy_r\int\frac{d^4q}{(2\pi)^4} \nonumber \\
&\times &\Big[ K^{RL}(p,q;Q) + K^{t}(p,q;Q) + K^{s}(p,q,r;Q)  \nonumber \\
 &+&K^{u}(p,q,r;Q) \Big]\left[S(q_1)\Psi(q;Q)S(q_2)\right]~,
\end{eqnarray}
with the momenta parametrised (in the rest frame of the bound state) as
\begin{eqnarray}
Q &=& (0, 0 , 0 , sqrt{Q^2})  \nonumber \\
p &=& \sqrt{p^2} ( 0,0,\sqrt{1 - z_p^2}, z_p) \nonumber \\
q &=& \sqrt{q^2} ( \sqrt{1 - z_q^2}\sqrt{1 - y_q^2} \sin{\phi_q}, \nonumber \\ &&\sqrt{1 - z_q^2}\sqrt{1 - y_q^2} \cos{\phi_q},\sqrt{1 - z_q^2} y_q, z_q) \nonumber \\
r &=& \sqrt{r^2} ( \sqrt{1 - z_r^2}\sqrt{1 - y_r^2} \sin{\phi_r}, \nonumber \\ &&\sqrt{1 - z_r^2}\sqrt{1 - y_r^2} \cos{\phi_r},\sqrt{1 - z_r^2} y_r, z_r)~.
\end{eqnarray}
As discussed in Sec.~\ref{subsec:branchcut}, after integrating over $z_r$, branch cuts originating from the pion and quark propagators are generated. A closed branch cut stemming from the pion propagators will overlap the positive-$r^2$ real axis when the total momentum $Q^2$ is real and negative and below the two-pion production threshold $Q^2<-4m_\pi^2$. To be able to perform the integration we must include an imaginary part to $Q^2$; in this way the cut opens and a deformation of the $r^2$ integration path can be performed to solve the BSE. 

We studied different parametrisations of the $r^2$ integration contour that avoid the branch cuts. For the results showed in this paper we used an integration path for $r^2$ formed by the union of two parametric segments
\begin{eqnarray}
r_1(t_1) &=& -\frac{Q^2_{real}}{a} t_1 \cos{t_1} - i \frac{Q^2_{real}}{a} t_1 \sin{t_1} \\
r_2(t_2) &=& -\frac{Q^2_{real}}{a} 2 \pi  + t_2~,\label{eq:contour_definition}
\end{eqnarray}
with $t_1 = [0, 2 \pi]$, $t_2 = [ 0 , UV + \frac{Q^2_{real}}{a} 2 \pi ]$ and $Q^2_{real}$ is the real part of $Q^2=-M^2+i\Gamma~M$ . We choose this paremetrisation for the real and imaginary parts of $Q^2$ since, when solving a homogeneous BSE, they would represent the mass $M$ and width $\Gamma$ of a solution with total momentum $Q$; for the calculation of the vertex herein, they should just be interpreted as the real and imaginary parts of the photon momentum. The contour parameter $a$ has to be adjusted for different values of the real and imaginary parts of $Q^2$ in order to avoid the different branch cuts. The values of $a$ for some representative choices of $M$ and $\Gamma$ are given in Table \ref{table:parameter_a}. Note that in some cases with small width no contour deformation is needed since there is not overlap of the cuts with the real axis. 

\begin{table}[h]
\begin{center}
\begin{tabular}{ |c|c|c|c|c| }   
\hline
  $M$  & $\Gamma$ & $a$ \\
\hline
0.500  & 0.001 to 0.400  & 6.0\\  
\hline
0.450  & 0.001 to 0.500  & 5.0\\  
  	   & 0.500 to 0.600  & 4.0\\  
\hline
0.400  & 0.001 to 0.400  & 6.0\\  
	 & 0.400 to 0.600  & 4.0\\  
\hline
0.350  & 0.001 to 0.100  & 8.2\\  
 & 0.100 to 0.600  & 4.0\\  
\hline
0.300  & 0.001 to 0.100  & 20.0\\  
 & 0.100 to 0.600  & 4.0\\  
\hline
0.276  & 0.001 to 0.010  & 600.0\\  
 & 0.010 to 0.100  & 60.0\\  
 & 0.100 to 0.600  & 5.0\\ 
\hline
0.250 to 0.275  & 0.001 to 0.010  & $\ast$\\  
   & 0.010 to 0.100  & 40.0\\  
    & 0.100 to 0.600  & 4.0\\    
\hline
0.200 to 0.250  & 0.001 to 0.200  & $\ast$\\ 

  & 0.200 to 0.600  & 4.0\\  
\hline
0.010 to 0.200 & 0.001 to 0.500  & $\ast$\\  
\hline
\end{tabular}
\end{center}
\caption{Different values for the parameter $a$ in Eqs.\eqref{eq:contour_definition}. In some cases no contour deformation is needed, these cases are represented by $\ast$.}
\label{table:parameter_a}
\end{table}


\section{Basis for the quark-photon vertex}
\label{sec:appendix_basis}

For the numerical calculation of the quark-photon vertex it is convenient to use the following basis
\begin{flalign}\label{eq:orthonormal_qphv_basis}
T_1\left(Q,p\right)&=\frac{1}{2\sqrt{2}}\gamma_{TT}^\mu~,
&T_2\left(Q,p\right)&=\frac{1}{2\sqrt{2}}\gamma_{TT}^\mu\slashed{\widehat{Q}}~,\nonumber\\
T_3\left(Q,p\right)&=\frac{1}{2\sqrt{2}}\gamma_{TT}^\mu\slashed{\widehat{p_T}}~,
&T_4\left(Q,p\right)&=\frac{1}{2\sqrt{2}}\gamma_{TT}^\mu\slashed{\widehat{p_T}}\slashed{\widehat{Q}}~,\nonumber\\
T_5\left(Q,p\right)&=\frac{1}{2}\widehat{p_T}^\mu~,
&T_6\left(Q,p\right)&=\frac{1}{2}\widehat{p_T}^\mu\slashed{\widehat{Q}}~,\nonumber\\
T_7\left(Q,p\right)&=\frac{1}{2}\widehat{p_T}^\mu\slashed{\widehat{p_T}}~,
&T_8\left(Q,p\right)&=\frac{1}{2}\widehat{p_T}^\mu\slashed{\widehat{p_T}}\slashed{\widehat{Q}}~,\nonumber\\
T_9\left(Q,p\right)&=\frac{i}{2}\widehat{Q}^\mu~,
&T_{10}\left(Q,p\right)&=\frac{1}{2}\widehat{Q}^\mu\slashed{\widehat{Q}}~,\nonumber\\
T_{11}\left(Q,p\right)&=\frac{1}{2}\widehat{Q}^\mu\slashed{\widehat{p_T}}~,
&T_{12}\left(Q,p\right)&=\frac{i}{2}\widehat{Q}^\mu\slashed{\widehat{p_T}}\slashed{\widehat{Q}}~,\nonumber
\end{flalign}
where we defined 
\begin{flalign}
 \gamma_{TT}^\mu&=\left(\delta^{\mu\nu}-\frac{p_T^\mu p_T^\nu}{p_T^2}\right)\left(\delta^{\nu\rho}-\frac{Q^\nu Q^\rho}{Q^2}\right)\gamma^\rho~,\\
 p_T^\mu&=\left(\delta^{\mu\nu}-\frac{Q^\mu Q^\nu}{Q^2}\right)p^\nu~,
\end{flalign}
and vectors with hat are normalized.

The following elements project onto each of the elements of the basis
\begin{flalign}\label{eq:projector_orthonormal_qphv_basis}
\bar{T}_1\left(Q,p\right)&=\frac{1}{2\sqrt{2}}\gamma_{TT}^\mu~,
&\bar{T}_2\left(Q,p\right)&=\frac{-1}{2\sqrt{2}}\gamma_{TT}^\mu\slashed{\widehat{Q}}~,\nonumber\\
\bar{T}_3\left(Q,p\right)&=\frac{-1}{2\sqrt{2}}\gamma_{TT}^\mu\slashed{\widehat{p_T}}~,
&\bar{T}_4\left(Q,p\right)&=\frac{-1}{2\sqrt{2}}\gamma_{TT}^\mu\slashed{\widehat{p_T}}\slashed{\widehat{Q}}~,\nonumber\\
\bar{T}_5\left(Q,p\right)&=\frac{1}{2}\widehat{p_T}^\mu~,
&\bar{T}_6\left(Q,p\right)&=\frac{1}{2}\widehat{p_T}^\mu\slashed{\widehat{Q}}~,\nonumber\\
\bar{T}_7\left(Q,p\right)&=\frac{1}{2}\widehat{p_T}^\mu\slashed{\widehat{p_T}}~,
&\bar{T}_8\left(Q,p\right)&=\frac{-1}{2}\widehat{p_T}^\mu\slashed{\widehat{p_T}}\slashed{\widehat{Q}}~,\nonumber\\
\bar{T}_9\left(Q,p\right)&=\frac{-i}{2}\widehat{Q}^\mu~,
&\bar{T}_{10}\left(Q,p\right)&=\frac{1}{2}\widehat{Q}^\mu\slashed{\widehat{Q}}~,\nonumber\\
\bar{T}_{11}\left(Q,p\right)&=\frac{1}{2}\widehat{Q}^\mu\slashed{\widehat{p_T}}~,
&\bar{T}_{12}\left(Q,p\right)&=\frac{i}{2}\widehat{Q}^\mu\slashed{\widehat{p_T}}\slashed{\widehat{Q}}~,\nonumber
\end{flalign}
in the sense that
\begin{flalign}
 \textrm{Tr}\left[\bar{T}_iT_j\right]&=\delta_{ij}~.
\end{flalign}

To plot our results we used instead the basis in Eqs.\eqref{eq:vertex_expansion1} and \eqref{eq:vertex_expansion2}. In order to rotate the dressings obtained for the basis \eqref{eq:orthonormal_qphv_basis} onto \eqref{eq:vertex_expansion1} and \eqref{eq:vertex_expansion2} we used the projectors\hfill
\flaligne{
P_1=&((-(p\cdot Q\slashed{p})+p^2~\slashed{Q})~Q^\mu)/(-4~(p\cdot Q)^2+4~p^2~Q^2)\nonumber\\
          P_2=&((-(Q^2\slashed{p})+p\cdot Q~\slashed{Q})~Q^\mu)/(8~((p\cdot Q)^3-p^2~p\cdot Q~Q^2))\nonumber\\
          P_3=&((i/8)~Q^\mu)/p\cdot Q\nonumber\\
          P_4=&((i/16)~((\slashed{p}\slashed{Q})-(\slashed{Q}\slashed{p}))~Q^\mu)/ 
                                    ((p\cdot Q)^2-p^2~Q^2)\nonumber\\
          P_5=&(p^2~Q^2~(p^2~\gamma^\mu-\slashed{p}~p^\mu)+3p^2~\slashed{Q}~(p\cdot Q~p^\mu-p^2~Q^\mu)+  \nonumber\\
                            &   p\cdot Q~(-(p\cdot Q~(p^2~\gamma^\mu+2\slashed{p}~p^\mu))+3p^2~\slashed{p}~Q^\mu))/ \nonumber\\
                            &                (8((p\cdot Q)^2-p^2~Q^2)^2)\nonumber\\
          P_6=&((i/16)~(-(((p\cdot Q)^2-p^2~Q^2)~(-\slashed{p}\cdot \gamma^\mu)+ \nonumber\\
                            &       (\gamma^\mu\cdot\slashed{p})))+(\slashed{Q}\slashed{p})~(3p\cdot Q~p^\mu-3p^2~Q^\mu)+ \nonumber\\
                            &       3(\slashed{p}\slashed{Q})(-(p\cdot Q~p^\mu)+p^2~Q^\mu)))/(p\cdot Q~((p\cdot Q)^2-p^2~Q^2)^2)\nonumber
\\
          P_7=&((i/16)~(p^\mu ((\slashed{p}\slashed{Q})-(\slashed{Q}\slashed{p}))+ \nonumber \\
                            &    p\cdot Q ~(-(\slashed{p}\gamma^\mu)+(\gamma^\mu\cdot\slashed{p}))+ \nonumber \\
                            &    p^2~((\slashed{Q}\gamma^\mu)-(\gamma^\mu\cdot\slashed{Q}))))/((p\cdot Q)^2-p^2~Q^2)\nonumber\\
          P_8=&(((\slashed{p}\slashed{Q})\cdot\gamma^\mu)-((\slashed{p}\gamma^\mu)\cdot\slashed{Q})-  ((\slashed{Q}\slashed{p})\cdot\gamma^\mu)+  \nonumber\\
                            &  ((\slashed{Q}\gamma^\mu)\cdot\slashed{p})+ ((\gamma^\mu\cdot\slashed{p})\cdot\slashed{Q})-   ((\gamma^\mu\cdot\slashed{Q})\cdot\slashed{p}))/ \nonumber\\
                            & (48(p\cdot Q)^2-48p^2~Q^2)\nonumber\\
          P_9=&((i/4)~(p\cdot Q~p^\mu-p^2~Q^\mu))/((p\cdot Q)^3-p^2~p\cdot Q~Q^2) \nonumber\\
          P_{10}=&(-(Q^2~(p\cdot Q~(p^2~\gamma^\mu-3\slashed{p}~p^\mu)+2p^2~\slashed{p}~Q^\mu))+  \nonumber\\
                            &   p\cdot Q~((p\cdot Q)^2~\gamma^\mu+3p^2~\slashed{Q}~Q^\mu-  \nonumber\\
                            &   p\cdot Q~(3\slashed{Q}~p^\mu+\slashed{p}~Q^\mu)))/(8p\cdot Q~((p\cdot Q)^2-p^2~Q^2)^2)  \nonumber\\
          P_{11}=&(\slashed{Q}~(-(((p\cdot Q)^2+2p^2~Q^2)~p^\mu)+ \nonumber\\
                            &   3p^2~p\cdot Q~Q^\mu)+p\cdot Q~(((p\cdot Q)^2-p^2~Q^2)~\gamma^\mu+ \nonumber\\
                            &   \slashed{p}~(3Q^2~p^\mu-3p\cdot Q~Q^\mu)))/(8p\cdot Q~((p\cdot Q)^2-p^2~Q^2)^2)\nonumber
}
\flaligne{
          P_{12}=&((i/16)~(((p\cdot Q)^2-p^2~Q^2)~(-(\slashed{Q}\gamma^\mu)+ \nonumber\\
                            &    (\gamma^\mu\cdot\slashed{Q}))+(\slashed{p}\cdot \slashed{Q})\times \nonumber\\
                            &    (3Q^2~p^\mu-3p\cdot Q~Q^\mu)+  \nonumber\\
                            &    3(\slashed{Q}\slashed{p})~(-(Q^2~p^\mu)+p\cdot Q~Q^\mu)))/((p\cdot Q)^2-p^2~Q^2)^2.\nonumber\\
\label{eq:projectors_EichmannBasis}}
We have 
\flaligne{
 \textrm{Tr}\left[P_i B_j\right]&=\delta_{ij}~,
}
with $B_i\equiv\{\gamma^\mu,2p^\mu\slashed{p},-2ip^\mu,i[\gamma^\mu,\slashed{p}],\tau_{1\dots 8}\}$ and $\tau_i$ defined in \eqref{eq:vertex_expansion2}.

\section{Poles of the quark propagator}\label{sec:quark_poles}

In the truncations of the quark DSE used in this paper, the quark propagator features pairs of complex conjugate poles in the complex plane (see, e.g. \cite{Windisch:2016iud}). In order to define an integration contour as in Appendix.~\ref{sec:contour_deformation} that avoids also the cuts generated by those poles, it is useful to parametrise the quark propagator $S(p) = -i \slashed{p} \sigma_v(p^2) + \sigma_s(p^2)$ simply as a sum of poles (see e.g. \cite{El-Bennich:2016qmb})
\begin{eqnarray}
\sigma_v &=& \sum_{i}^n \left[\frac{\alpha_i}{p^2 + m_i} + \frac{\alpha_i^\ast}{p^2 + m_i^\ast}\right] \nonumber \\ \sigma_s &=& \sum_{i}^n \left[\frac{\beta_i}{p^2 + m_i} + \frac{\beta_i^\ast}{p^2 + m_i^\ast}\right]~,
\end{eqnarray}
where the parameters $m_i$, $\alpha_i$, $\beta_i$ can be obtained by fitting the corresponding quark DSE solution along the $p^2$ real axis or, alternatively, on a  parabola in the complex plane that does not enclose the poles. Our fitted values for the pole positions $m_i$ are given in Table \ref{table:poles} where for the fits we used two pairs of complex conjugate poles. 
\begin{table*}[h]
\begin{center}
\begin{tabular}{|l|l|l|l|l|}
\hline
       & \multicolumn{2}{l|}{Real axis} & \multicolumn{2}{l|}{Parabola} \\ \hline
       & R-L            & P-E           & R-L           & P-E           \\ \hline
$m_1$ &  0.2549 $\pm$ 0.0481  &  0.2081 $\pm$ 0.0596             &     0.2560 $\pm$ 0.0486       &   0.2448 $\pm$ 0.0516           \\
\hline
$m_2$ &   1.8982 $\pm$ 1.8687    &  0.3833 $\pm$ 0.8158             & 1.1881 $\pm$ 0.0775       &    2.2316 $\pm$ 0.0510 \\
\hline
\end{tabular}
\end{center}
\caption{Complex conjugate poles fitting along the real axis and on a parabola in the $p^2$ complex plane, with apex at $p^2=-0.16$. Only the first pole is stable, the second strongly depends on the method used for the fit.}
\label{table:poles}
\end{table*}
Note, however, that we only used this ansatz for the quark propagator in order to find the parameter $a$ in the deformed contour \eqref{eq:contour_definition} and to determine the cuts in Fig.~\ref{Fig:cutspoles}. For the calculation of the quark-photon vertex we used a numerical solution of the quark DSE.  

%

\begin{thebibliography}{99}

\bibitem{Aznauryan:2012ba}
  I.~G.~Aznauryan {\it et al.},
  Int.\ J.\ Mod.\ Phys.\ E {\bf 22} (2013) 1330015
  doi:10.1142/S0218301313300154
  [arXiv:1212.4891 [nucl-th]].



\bibitem{Asner:2008nq}
  D.~M.~Asner {\it et al.},
  Int.\ J.\ Mod.\ Phys.\ A {\bf 24} (2009) S1
  [arXiv:0809.1869 [hep-ex]].



\bibitem{Wiedner:2011mf}
  U.~Wiedner,
  Prog.\ Part.\ Nucl.\ Phys.\  {\bf 66} (2011) 477
  doi:10.1016/j.ppnp.2011.04.001
  [arXiv:1104.3961 [hep-ex]].



\bibitem{Denig:2012by}
  A.~Denig and G.~Salme,
  Prog.\ Part.\ Nucl.\ Phys.\  {\bf 68} (2013) 113
  doi:10.1016/j.ppnp.2012.09.005
  [arXiv:1210.4689 [hep-ex]].



\bibitem{Pacetti:2015iqa}
  S.~Pacetti, R.~Baldini Ferroli and E.~Tomasi-Gustafsson,
  Phys.\ Rept.\  {\bf 550-551} (2015) 1.
  doi:10.1016/j.physrep.2014.09.005



\bibitem{Punjabi:2015bba}
  V.~Punjabi, C.~F.~Perdrisat, M.~K.~Jones, E.~J.~Brash and C.~E.~Carlson,
  Eur.\ Phys.\ J.\ A {\bf 51} (2015) 79
  doi:10.1140/epja/i2015-15079-x
  [arXiv:1503.01452 [nucl-ex]].



\bibitem{Ball:1980ay}
  J.~S.~Ball and T.~W.~Chiu,
  Phys.\ Rev.\ D {\bf 22} (1980) 2542.
  doi:10.1103/PhysRevD.22.2542

\bibitem{Frank:1994mf}
  M.~R.~Frank,
  Phys.\ Rev.\ C {\bf 51} (1995) 987
  doi:10.1103/PhysRevC.51.987
  [nucl-th/9403009].


\bibitem{Maris:1999bh}
  P.~Maris and P.~C.~Tandy,
  Phys.\ Rev.\ C {\bf 61} (2000) 045202
  doi:10.1103/PhysRevC.61.045202
  [nucl-th/9910033].



\bibitem{Chang:2010hb}
  L.~Chang, Y.~X.~Liu and C.~D.~Roberts,
  Phys.\ Rev.\ Lett.\  {\bf 106} (2011) 072001
  doi:10.1103/PhysRevLett.106.072001
  [arXiv:1009.3458 [nucl-th]].



\bibitem{Eichmann:2014qva}
  G.~Eichmann,
  Acta Phys.\ Polon.\ Supp.\  {\bf 7} (2014) no.3,  597
  doi:10.5506/APhysPolBSupp.7.597
  [arXiv:1404.4149 [nucl-th]].



\bibitem{Tang:2019zbk}
  C.~Tang, F.~Gao and Y.~X.~Liu,
  arXiv:1902.01679 [hep-ph].

\bibitem{Sakurai}
	J.~J.~Sakurai,
	Ann.\ Phys.\ {\bf 11} (1960)

\bibitem{Bauer:1977iq}
  T.~H.~Bauer, R.~D.~Spital, D.~R.~Yennie and F.~M.~Pipkin,
  Rev.\ Mod.\ Phys.\  {\bf 50} (1978) 261
   Erratum: [Rev.\ Mod.\ Phys.\  {\bf 51} (1979) 407].
  doi:10.1103/RevModPhys.50.261



\bibitem{Leupold:2012qn}
  S.~Leupold and C.~Terschlusen,
  PoS BORMIO {\bf 2012} (2012) 024
  doi:10.22323/1.160.0024
  [arXiv:1206.2253 [hep-ph]].



\bibitem{Cloet:2013jya}
  I.~C.~Cloet and C.~D.~Roberts,
  Prog.\ Part.\ Nucl.\ Phys.\  {\bf 77} (2014) 1
  doi:10.1016/j.ppnp.2014.02.001
  [arXiv:1310.2651 [nucl-th]].



\bibitem{Eichmann:2016yit}
  G.~Eichmann, H.~Sanchis-Alepuz, R.~Williams, R.~Alkofer and C.~S.~Fischer,
  Prog.\ Part.\ Nucl.\ Phys.\  {\bf 91} (2016) 1
  doi:10.1016/j.ppnp.2016.07.001
  [arXiv:1606.09602 [hep-ph]].

\bibitem{Huber:2018ned}
  M.~Q.~Huber,
  arXiv:1808.05227 [hep-ph].


\bibitem{Sanchis-Alepuz:2014wea}
  H.~Sanchis-Alepuz, C.~S.~Fischer and S.~Kubrak,
  Phys.\ Lett.\ B {\bf 733} (2014) 151
  doi:10.1016/j.physletb.2014.04.031
  [arXiv:1401.3183 [hep-ph]].



\bibitem{Williams:2015cvx}
  R.~Williams, C.~S.~Fischer and W.~Heupel,
  Phys.\ Rev.\ D {\bf 93} (2016) no.3,  034026
  doi:10.1103/PhysRevD.93.034026
  [arXiv:1512.00455 [hep-ph]].



\bibitem{Qin:2019hgk}
  S.~x.~Qin, C.~D.~Roberts and S.~M.~Schmidt,
  Few Body Syst.\  {\bf 60} (2019) no.2,  26
  doi:10.1007/s00601-019-1488-x
  [arXiv:1902.00026 [nucl-th]].



\bibitem{Maris:2000sk}
  P.~Maris and P.~C.~Tandy,
  Phys.\ Rev.\ C {\bf 62} (2000) 055204
  doi:10.1103/PhysRevC.62.055204
  [nucl-th/0005015].



\bibitem{Maris:2002mz}
  P.~Maris and P.~C.~Tandy,
  Phys.\ Rev.\ C {\bf 65} (2002) 045211
  doi:10.1103/PhysRevC.65.045211
  [nucl-th/0201017].



\bibitem{Bhagwat:2006pu}
  M.~S.~Bhagwat and P.~Maris,
  Phys.\ Rev.\ C {\bf 77} (2008) 025203
  doi:10.1103/PhysRevC.77.025203
  [nucl-th/0612069].



\bibitem{Cloet:2008re}
  I.~C.~Cloet, G.~Eichmann, B.~El-Bennich, T.~Klahn and C.~D.~Roberts,
  Few Body Syst.\  {\bf 46} (2009) 1
  doi:10.1007/s00601-009-0015-x
  [arXiv:0812.0416 [nucl-th]].



\bibitem{Nicmorus:2010sd}
  D.~Nicmorus, G.~Eichmann and R.~Alkofer,
  Phys.\ Rev.\ D {\bf 82} (2010) 114017
  doi:10.1103/PhysRevD.82.114017
  [arXiv:1008.3184 [hep-ph]].



\bibitem{Eichmann:2011vu}
  G.~Eichmann,
  Phys.\ Rev.\ D {\bf 84} (2011) 014014
  doi:10.1103/PhysRevD.84.014014
  [arXiv:1104.4505 [hep-ph]].



\bibitem{Eichmann:2011pv}
  G.~Eichmann and C.~S.~Fischer,
  Eur.\ Phys.\ J.\ A {\bf 48} (2012) 9
  doi:10.1140/epja/i2012-12009-6
  [arXiv:1111.2614 [hep-ph]].



\bibitem{Eichmann:2011aa}
  G.~Eichmann and D.~Nicmorus,
  Phys.\ Rev.\ D {\bf 85} (2012) 093004
  doi:10.1103/PhysRevD.85.093004
  [arXiv:1112.2232 [hep-ph]].



\bibitem{Sanchis-Alepuz:2013iia}
  H.~Sanchis-Alepuz, R.~Williams and R.~Alkofer,
  Phys.\ Rev.\ D {\bf 87} (2013) no.9,  096015
  doi:10.1103/PhysRevD.87.096015
  [arXiv:1302.6048 [hep-ph]].



\bibitem{Chang:2013nia}
  L.~Chang, I.~C.~Cloët, C.~D.~Roberts, S.~M.~Schmidt and P.~C.~Tandy,
  Phys.\ Rev.\ Lett.\  {\bf 111} (2013) no.14,  141802
  doi:10.1103/PhysRevLett.111.141802
  [arXiv:1307.0026 [nucl-th]].



\bibitem{Segovia:2014aza}
  J.~Segovia, I.~C.~Cloet, C.~D.~Roberts and S.~M.~Schmidt,
  Few Body Syst.\  {\bf 55} (2014) 1185
  doi:10.1007/s00601-014-0907-2
  [arXiv:1408.2919 [nucl-th]].



\bibitem{Sanchis-Alepuz:2015fcg}
  H.~Sanchis-Alepuz and C.~S.~Fischer,
  Eur.\ Phys.\ J.\ A {\bf 52} (2016) no.2,  34
  doi:10.1140/epja/i2016-16034-1
  [arXiv:1512.00833 [hep-ph]].



\bibitem{Segovia:2016zyc}
  J.~Segovia and C.~D.~Roberts,
  Phys.\ Rev.\ C {\bf 94} (2016) no.4,  042201
  doi:10.1103/PhysRevC.94.042201
  [arXiv:1607.04405 [nucl-th]].



\bibitem{Sanchis-Alepuz:2017mir}
  H.~Sanchis-Alepuz, R.~Alkofer and C.~S.~Fischer,
  Eur.\ Phys.\ J.\ A {\bf 54} (2018) no.3,  41
  doi:10.1140/epja/i2018-12465-x
  [arXiv:1707.08463 [hep-ph]].



\bibitem{Chen:2018nsg}
  C.~Chen, Y.~Lu, D.~Binosi, C.~D.~Roberts, J.~Rodríguez-Quintero and J.~Segovia,
  Phys.\ Rev.\ D {\bf 99} (2019) no.3,  034013
  doi:10.1103/PhysRevD.99.034013
  [arXiv:1811.08440 [nucl-th]].



\bibitem{Chen:2018rwz}
  M.~Chen, M.~Ding, L.~Chang and C.~D.~Roberts,
  Phys.\ Rev.\ D {\bf 98} (2018) no.9,  091505
  doi:10.1103/PhysRevD.98.091505
  [arXiv:1808.09461 [nucl-th]].



\bibitem{Ding:2018xwy}
  M.~Ding, K.~Raya, A.~Bashir, D.~Binosi, L.~Chang, M.~Chen and C.~D.~Roberts,
  Phys.\ Rev.\ D {\bf 99} (2019) no.1,  014014
  doi:10.1103/PhysRevD.99.014014
  [arXiv:1810.12313 [nucl-th]].



\bibitem{Fischer:2007ze}
  C.~S.~Fischer, D.~Nickel and J.~Wambach,
  Phys.\ Rev.\ D {\bf 76} (2007) 094009
  doi:10.1103/PhysRevD.76.094009
  [arXiv:0705.4407 [hep-ph]].



\bibitem{Fischer:2008sp}
  C.~S.~Fischer, D.~Nickel and R.~Williams,
  Eur.\ Phys.\ J.\ C {\bf 60} (2009) 47
  doi:10.1140/epjc/s10052-008-0821-1
  [arXiv:0807.3486 [hep-ph]].

\bibitem{Windisch:2012sz}
  A.~Windisch, M.~Q.~Huber and R.~Alkofer,
  Phys.\ Rev.\ D {\bf 87} (2013) no.6,  065005
  doi:10.1103/PhysRevD.87.065005
  [arXiv:1212.2175 [hep-ph]].

  
\bibitem{Windisch:2013dxa}
  A.~Windisch, M.~Q.~Huber and R.~Alkofer,
  Acta Phys.\ Polon.\ Supp.\  {\bf 6} (2013) no.3,  887
  doi:10.5506/APhysPolBSupp.6.887
  [arXiv:1304.3642 [hep-ph]].

\bibitem{Eichmann:2017wil}
  G.~Eichmann, C.~S.~Fischer, E.~Weil and R.~Williams,
  Phys.\ Lett.\ B {\bf 774} (2017) 425
  doi:10.1016/j.physletb.2017.10.002
  [arXiv:1704.05774 [hep-ph]].



\bibitem{Weil:2017knt}
  E.~Weil, G.~Eichmann, C.~S.~Fischer and R.~Williams,
  Phys.\ Rev.\ D {\bf 96} (2017) no.1,  014021
  doi:10.1103/PhysRevD.96.014021
  [arXiv:1704.06046 [hep-ph]].



\bibitem{Sanchis-Alepuz:2017jjd}
  H.~Sanchis-Alepuz and R.~Williams,
  Comput.\ Phys.\ Commun.\  {\bf 232} (2018) 1
  doi:10.1016/j.cpc.2018.05.020
  [arXiv:1710.04903 [hep-ph]].



\bibitem{Maris:1997tm}
  P.~Maris and C.~D.~Roberts,
  Phys.\ Rev.\ C {\bf 56} (1997) 3369
  doi:10.1103/PhysRevC.56.3369
  [nucl-th/9708029].



\bibitem{Maris:1999nt}
  P.~Maris and P.~C.~Tandy,
  Phys.\ Rev.\ C {\bf 60} (1999) 055214
  doi:10.1103/PhysRevC.60.055214
  [nucl-th/9905056].



\bibitem{Williams:2018adr}
  R.~Williams,
  arXiv:1804.11161 [hep-ph].

\bibitem{Windisch:2016iud}
  A.~Windisch,
  Phys.\ Rev.\ C {\bf 95} (2017) no.4,  045204
  doi:10.1103/PhysRevC.95.045204
  [arXiv:1612.06002 [hep-ph]].



\bibitem{El-Bennich:2016qmb}
  B.~El-Bennich, G.~Krein, E.~Rojas and F.~E.~Serna,
  Few Body Syst.\  {\bf 57} (2016) no.10,  955
  doi:10.1007/s00601-016-1133-x
  [arXiv:1602.06761 [nucl-th]].

\end{thebibliography}

\end{document}